\documentclass[10pt,twocolumn,letterpaper]{article}

\usepackage[pagenumbers]{cvpr} % To force page numbers, e.g. for an arXiv version

\usepackage[dvipsnames]{xcolor}

\definecolor{cvprblue}{rgb}{0.21,0.49,0.74}
\usepackage[pagebackref,breaklinks,colorlinks,citecolor=cvprblue]{hyperref}

\usepackage{graphicx}
\usepackage{amsmath}
\usepackage{amssymb}
\usepackage{booktabs}
\usepackage{kotex}  % pdfLaTeX korean supports
\usepackage{array}
\usepackage{tabularx}
\usepackage{multirow}
\usepackage{subcaption}
\usepackage[dvipsnames]{xcolor}
\usepackage{breakcites} % break cites
\usepackage{comment}
\usepackage{url}
\usepackage{colortbl}
\usepackage{nccmath}
\usepackage{makecell}
\usepackage[accsupp]{axessibility}

\newcommand{\parhead}[1]{\vspace{8pt}\noindent\textbf{#1}}

\newcommand{\fst}[1]{\textcolor[rgb]{0,0,0}{\textbf{#1}}}
\newcommand{\snd}[1]{\textcolor[rgb]{0,0,1}{#1}}

\newcommand{\capred}[1]{\textcolor{Red}{#1}}
\renewcommand{\paragraph}[1]{{\vspace{2pt}\noindent\textbf{#1}}}

\title{ParamISP: Learned Forward and Inverse ISPs using Camera Parameters}

\author{
Woohyeok Kim$^1$$^{\ast}$ ~ ~ ~
Geonu Kim$^1$$^{\ast}$ ~ ~ ~
Junyong Lee$^2$$^{\dagger}$ \\ [1mm] 
Seungyong Lee$^1$ ~ ~ ~
Seung-Hwan Baek$^1$ ~ ~ ~
Sunghyun Cho$^1$ \\ [2mm]
$^1$POSTECH ~ ~ ~ 
$^2$Samsung AI Center Toronto \\
}

\newcommand\blfootnote[1]{
  \begingroup
  \renewcommand\thefootnote{}\footnote{#1}
  \addtocounter{footnote}{-1}
  \endgroup
}

\begin{document}
\maketitle
\begin{abstract}
RAW images are rarely shared mainly due to its excessive data size compared to their sRGB counterparts obtained by camera ISPs. Learning the forward and inverse processes of camera ISPs has been recently demonstrated, enabling physically-meaningful RAW-level image processing on input sRGB images. 
However, existing learning-based ISP methods fail to handle the large variations in the ISP processes with respect to camera parameters such as ISO and exposure time, and have limitations when used for various applications. 
In this paper, we propose ParamISP, a learning-based method for forward and inverse conversion between sRGB and RAW images, that adopts a novel neural-network module to utilize camera parameters, which is dubbed as ParamNet.
Given the camera parameters provided in the EXIF data, ParamNet converts them into a feature vector to control the ISP networks.
Extensive experiments demonstrate that ParamISP achieve superior RAW and sRGB reconstruction results compared to previous methods and it can be effectively used for a variety of applications such as deblurring dataset synthesis, raw deblurring, HDR reconstruction, and camera-to-camera transfer.
\blfootnote{$^{\ast}$ Equal contribution.}
\blfootnote{$^{\dagger}$ Work done prior to joining Samsung.}
\end{abstract}
    
\vspace{-0.3cm}
\section{Introduction}
\label{sec:intro}

A camera ISP converts a RAW image into a visually pleasing sRGB image, which is typically saved as a JPEG image.
A camera ISP performs a series of operations including defective pixel correction, denoising, lens shading correction, white balance, color filter array interpolation, color space conversion, tone reproduction, and non-linear contrast enhancement.
Detailed operations of camera ISPs are typically sealed to the public and vary from camera to camera.

Unlike sRGB images, RAW images provide physically-meaningful and interpretable information such as noise distributions as they have the linear relationship between image intensity and radiant energy incident on a camera sensor.
Such properties of RAW images have been exploited for denoising~\cite{abdelhamed2018high,mildenhall2018burst,schwartz2018deepisp}, HDR~\cite{Chen_2018_CVPR,liu2020single}, and super-resolution~\cite{zhang2019zoom, tang2022learning}, leading to superior quality than using only sRGB images.
However, a RAW image demands large memory to store due to the use of high-precision bits without any compression. 
Therefore, only sRGB images are often shared without their RAW counterparts, making it difficult to utilize the useful properties of RAW images. 

\begin{figure}[t]
  \centering
  \includegraphics[width=1.0\linewidth]{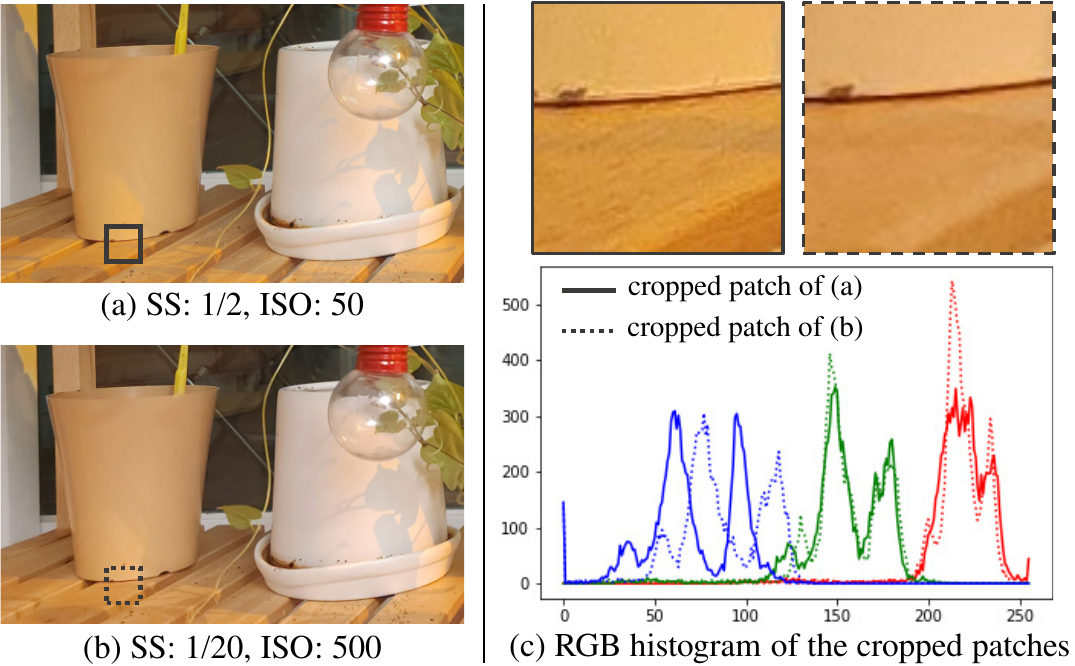}
  \vspace{-14pt}
  \caption{Impact of camera parameters on a camera ISP. Images (a) and (b) are taken by a Samsung Galaxy S22 with different camera parameters. SS and ISO indicate the shutter speed and sensor sensitivity, respectively. In (c), we visualize RGB histograms of each cropped patch of (a) and (b).
  Despite the same target scene, the captured images exhibit distinct histograms, implying complex ISP operations dependent on the camera parameters.}
  \vspace{-0.4cm}
  \label{fig:teas-isp_param}
\end{figure}

Recently, several approaches have been proposed to reconstruct RAW images from sRGB images and vice versa by modeling forward and inverse ISPs~\cite{brooks2019unprocessing,zamir2020cycleisp,xing2021invertible,afifi2021cie,conde2022model}.
However, despite notable improvements, they still suffer from limitations.
First, real-world ISPs adjust their operations based on the camera parameters, e.g., exposure time and sensor sensitivity, as shown in \cref{fig:teas-isp_param}.
However, previous methods overlook this adaptive nature of real-world ISPs, and learn average ISP operations, which leads to low reconstruction performance.
Second, previous methods adopt rather simple network architectures disregarding the complexity of the ISP operations, which leads to low RAW reconstruction quality.  
They construct a single network by stacking invertible or residual blocks~\cite{zamir2020cycleisp,xing2021invertible}, or organize modules by simply arranging convolution layers~\cite{afifi2021cie}.

In this paper, we present a novel forward and inverse ISP framework, \emph{ParamISP}.
ParamISP is designed to faithfully reflect the real-world ISP operations that change based on camera parameters.
To this end, ParamISP leverages camera parameters provided in the EXIF metadata of a JPEG image. 
To effectively incorporate the camera parameters, ParamISP consists of a pair of forward and inverse ISP networks that include a novel neural network module \emph{ParamNet}.
ParamNet extracts a feature vector from the camera parameters including exposure time, sensitivity, aperture size, and focal length, and feeds it to the forward and inverse ISP subnetworks to control their behaviors.

To learn ISP operations for varying camera parameters, we need to address the following problems.
First, the camera parameters have significantly different scales, such as an exposure time of 0.01 sec.~and a sensitivity of 800.
Moreover, some parameters such as exposure time and sensor sensitivity have non-linearly increasing parameter values, e.g., the pre-set exposure time of commodity cameras roughly increases exponentially (1/250 sec., 1/125 sec., 1/60 sec., ...).
Second, existing datasets lack of diversity in camera parameters~\cite{dang2015raise,schwartz2018deepisp,rim2020real}, while a significant effort is required to collect a sufficient number of images for possible combinations of different camera parameters.
Such scale difference and insufficient amount of training data make it difficult to reliably learn ISP operations for different camera parameters.
Therefore, we propose a non-linear equalization scheme that adjusts the scales of the camera parameters and a random-dropout-based learning strategy to effectively learn the effects of all the camera parameters.

Finally, we present novel network architectures that reflect the real-world ISP operations.
Specifically, our ISP networks consist of four subnetworks: \emph{CanoNet}, \emph{LocalNet} and \emph{GlobalNet}, along with \emph{ParamNet}.
CanoNet performs common ISP operations such as demosaicing, white balance, and color space conversion using fixed operations without learnable weights.
GlobalNet performs global tone manipulation with respect to the camera parameters.
LocalNet performs other residual operations that are not captured by CanoNet and GlobalNet.

By incorporating camera parameters, novel network architectures, and training schemes, ParamISP achieves accurate reconstruction performance with a smaller network size.
Our extensive evaluation shows that ParamISP surpasses previous learning-based ISP models by an average of 1.93 dB and 1.84 dB in RAW and sRGB reconstruction, respectively (\cref{sec:exp}).
Furthermore, ParamISP is robustly applicable to various applications (\eg, deblurring dataset synthesis, RAW deblurring, HDR reconstruction, camera-to-camera transfer) (\cref{sec:app}).

Our contributions are summarized as follows:
\begin{itemize}
\item[$\bullet$]
We propose \emph{ParamISP}, a novel learning-based forward and inverse ISP framework that leverages camera parameters (\eg, exposure time, sensitivity, aperture size, and focal length) for high-quality sRGB/RAW reconstruction.
\item[$\bullet$] 
To effectively incorporate camera parameters, we develop \emph{ParamNet}, a novel neural network module that controls the forward and inverse ISP networks according to camera parameters. We also introduce a non-linear equalization scheme and a random-dropout-based learning strategy for stable and effective learning of ISP operations with respect to camera parameters.
\item[$\bullet$]
We present novel network architectures for the forward and inverse ISP networks that better reflect real-world ISPs.
(CanoNet, LocalNet, GlobalNet, ParamNet)
\item[$\bullet$] We demonstrate the performance of \emph{ParamISP} and its internal modules on many cameras, 
and show its applicability to various applications.
\end{itemize}

\section{Related Work}
\label{sec:relwork}

\begin{figure*}[t!]
\centering
\includegraphics[width=0.95\linewidth]{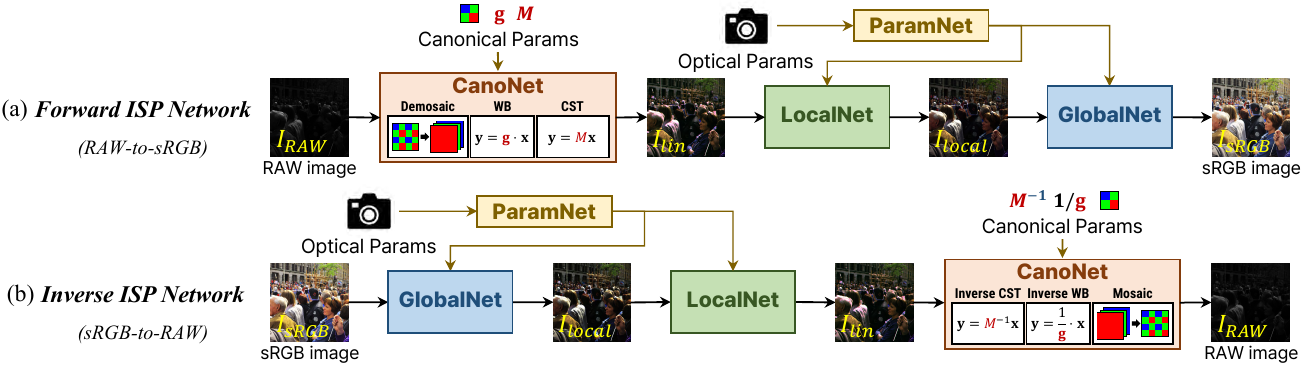}
\vspace{-0.2cm}
\caption{Overview of the proposed ParamISP framework. The full pipeline is constructed by combining learnable networks (ParamNet, LocalNet, GlobalNet) with invertible canonical camera operations (CanoNet). CanoNet consists of differentiable operations without learnable weights, where WB and CST denote white balance and color space transform, respectively.}
\vspace{-0.3cm}
\label{fig:overview}
\end{figure*}

\paragraph{Parametric ISPs}
To approximate the mapping between RAW and sRGB images, ISP approaches based on non-learnable parametric operations~\cite{brooks2019unprocessing,conde2022model} have been proposed.
They employ simple invertible ISPs that are composed of non-learnable operations using either camera parameters~\cite{brooks2019unprocessing} (\eg, white balance and color correction matrices) or parameters learned from DNNs~\cite{conde2022model}.
However, they do not consider complex nonlinear ISP operations (\eg, local tone mapping and denoising), resulting in inaccurate reconstruction of sRGB and RAW images.

\parhead{Learnable ISPs}
For more accurate approximation to forward and inverse ISP operations, DNN-based approaches~\cite{zamir2020cycleisp, xing2021invertible, afifi2021cie} have been proposed, for which symmetrical forward and inverse ISP networks~\cite{zamir2020cycleisp, afifi2021cie} and an invertible ISP network~\cite{xing2021invertible} are employed to learn mapping between sRGB and RAW images.
However, they are designed without considering camera parameters (\eg, exposure time, sensor sensitivity, etc), which limit their RAW and sRGB reconstruction quality.

Moreover, some of the previous methods primarily focus on cyclic reconstruction (sRGB-to-RAW-to-sRGB) and try to minimize the difference between the input sRGB image and the sRGB image restored back by the inverse and forward ISPs.
Specifically, CycleISP~\cite{zamir2020cycleisp} uses the input sRGB image for restoring the tone when reconstructing an sRGB image back from a RAW image, which makes it unsuitable for applications that manipulate the tone in the RAW space.
InvISP~\cite{xing2021invertible} employs a single normalizing flow-based invertible network for both inverse and forward processes.
While this design choice leads to near perfect reconstruction quality for cyclic reconstruction, its quality significantly degrades when the intermediate RAW images are altered, making it unsuitable for applications that manipulate images in the RAW space as demonstrated in \cref{sec:app}.
In contrast, we design our forward and inverse ISP networks to operate independently during inference time to cater to a broader range of applications.

\paragraph{RAW Reconstruction using Additional Information}
To achieve precise RAW reconstruction from an sRGB image, another line of research that utilizes additional information has been introduced~\cite{punnappurath2021spatially,nam2022learning,li2023metadata,wang2023raw}. This approach encodes necessary metadata such as a small portion of a RAW image within an sRGB image at capture time to reconstruct the original RAW image with high accuracy.
However, it necessitates a modification to the existing ISP process to store the metadata.
In contrast, our approach utilizes the EXIF metadata that commodity cameras already provide.

\section{ParamISP}
\label{sec:method}

Given a target camera, our goal is to learn its forward and inverse ISP processes that change with respect to camera parameters.
To accomplish this, ParamISP is designed to have a pair of forward (RAW-to-sRGB) and inverse (sRGB-to-RAW) ISP networks (\cref{fig:overview}).
Both networks are equipped with ParamNet so that they adaptively operate based on camera parameters.

In ParamISP, we classify camera parameters into two distinct categories: optical parameters (including exposure time, sensitivity, aperture size, and focal length) and canonical parameters (Bayer pattern, white balance coefficients, and a color correction matrix). 
The canonical parameters directly influence fundamental ISP operations like demosaicing, white balance adjustments, and color space conversion. 
These operations are relatively straightforward, and their relationship with the canonical parameters is well-defined. 

While previous approaches have leveraged the canonical parameters~\cite{brooks2019unprocessing,conde2022model}, the optical parameters have remained untapped. In contrast, ParamISP exploits both sets of parameters to achieve highly accurate sRGB and RAW reconstruction. 
To harness the canonical parameters, our ISP networks incorporate CanoNet, a subnetwork that performs canonical ISP operations without learnable weights. 
For the optical parameters, we introduce ParamNet, which is the key component to dynamically control the behavior of the ISP networks based on the optical parameters.

In the following, we describe ParamNet and our training strategy for stable and effective training.
We then explain the forward and inverse ISP subnetworks in detail.

\subsection{ParamNet}
\label{ssec:paramnet}

\begin{figure*}[t!]
\centering
\includegraphics[width=0.95\linewidth]{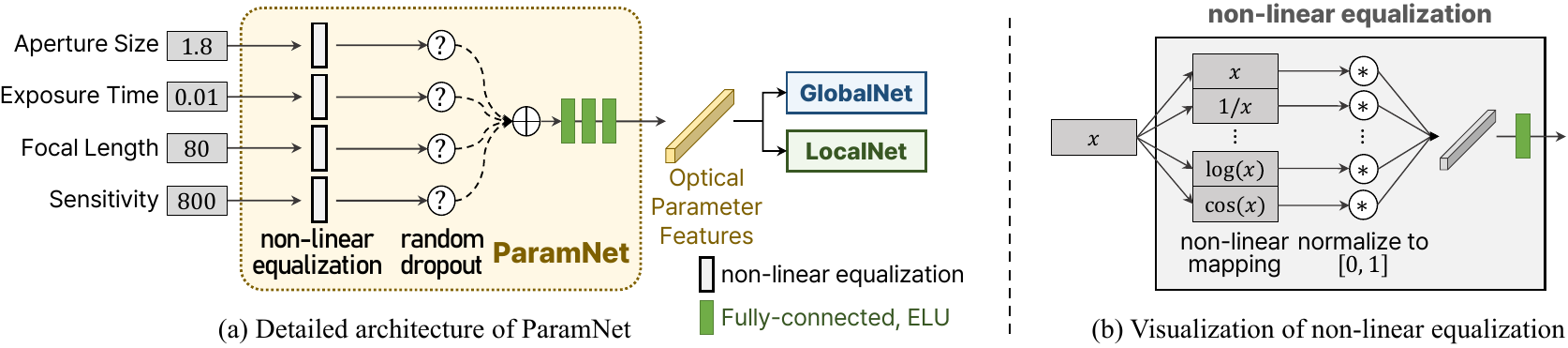}
\vspace{-0.2cm}
\caption{Architecture of ParamNet. (a) Given camera optical parameters, ParamNet estimates optical parameter features used for modulating the LocalNet and GlobalNet. (b) In order to deal with different scales and non-linearly distributed values of optical parameters, we propose to use non-linear equalization that exploits multiple non-linear mapping functions.}
\vspace{-0.3cm}
\label{fig:paramnet}
\end{figure*}

As described in \cref{fig:paramnet}, ParamNet takes optical parameters as input and converts them into an optical parameter feature vector $z$, which is then fed to both LocalNet and GlobalNet.
ParamNet consists of a non-linear equalization layer, and fully-connected layers.
The non-linear equalization layer computes a normalized feature vector for each optical parameter to compensate for the scale difference between the optical parameters.
The equalized feature vectors are then arithmetically summed together, and fed to the fully-connected layers to obtain an optical parameter feature vector $z$.

\paragraph{Non-linear Equalization}
As discussed in \cref{sec:intro}, the optical parameters exhibit significantly different scales, and their values are non-linearly distributed. 
While some optical parameters may substantially alter the behavior of the camera ISP, others may have minimal effect. Moreover, the influence of optical parameters may vary across different camera models.
As a result, na\"{i}vely incorporating the optical parameters leads to unstable training of ParamNet.

To resolve this issue, the non-linear equalization layer of ParamNet applies various non-linear mapping to each optical parameter and learns the best composition of them that enables stable training.
Specifically, we first apply non-linear mapping functions such as $x$, $1/x$, $\sqrt{x}$, $x^{-1/2}$, $x^{1/4}$, $x^{-1/4}$, $\log(x)$, $\sin(\log(x))$, $\cos(\log(x))$, $\sin(c\cdot x)$, and $\cos(c\cdot x)$ to each optical parameter,
where $x$ is the value of an optical parameter, and $c$ controls the frequency of the sinusoidal functions, for which we use three different values empirically chosen.
We then normalize each of the non-linear mapping results into $[0, 1]$.
As a result, we obtain a $15$-dim.~vector for each optical parameter.
Then, each vector is processed through a subsequent fully-connected layer to extract useful information from each parameter vector.

Our non-linear equalization layer incorporates a range of diverse non-linear mapping functions, rather than relying on a predefined set of carefully-chosen functions, in order to cover the wide spectrum of potential relationships between each optical parameter and the camera ISP.
Thus, some mapping functions may look redundant or unnecessary.
Nevertheless, the subsequent fully-connected layer can successfully learn to extract only useful information from them by combining them with different weights.

\paragraph{Random Dropout}
Even with the non-linear equalization, ParamNet can still be trained to overfit to a subset of the optical parameters due to the scale difference and the insufficient amount of training data.
To mitigate the problem, during training, we randomly drop out the equalized feature vector of each optical parameter.
Specifically, each optical parameter is randomly dropped out by the probablity of 20\% at each training iteration.

\subsection{Forward ISP Network}
\label{ssec:ispnetwork}

\begin{figure*}[t!]
\includegraphics[width=1.0\linewidth]{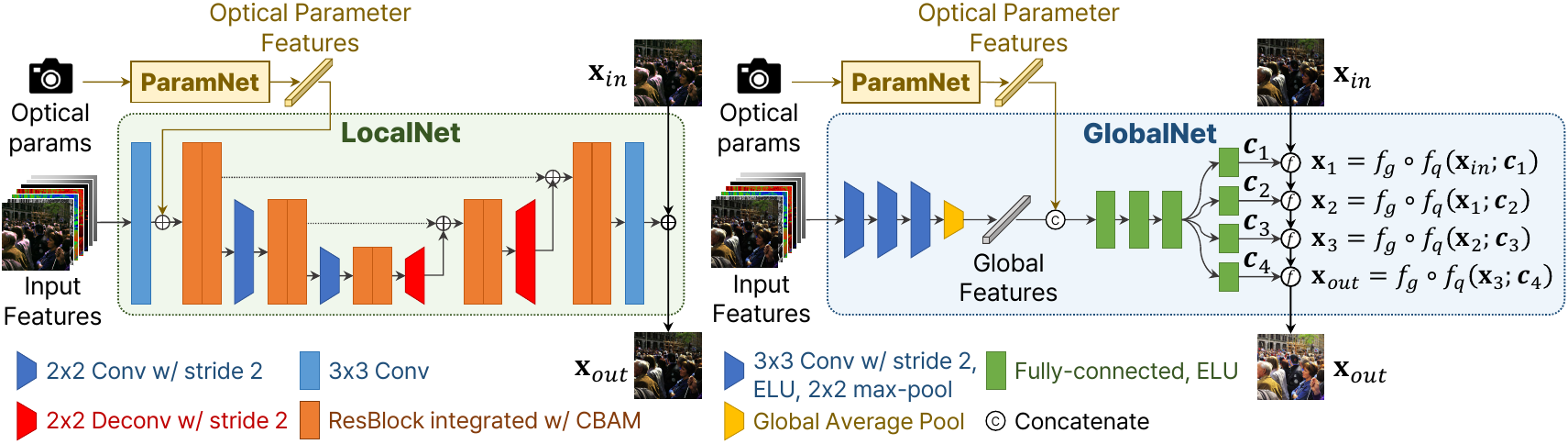}
\vspace{-0.6cm}
\caption{Detailed architecture of LocalNet and GlobalNet. In GlobalNet, $f_g$ and $f_q$ represent the gamma correction and quadratic transformation, respectively, while $c_n$ represents the $n$-th coefficients $G^n$ and $W^n$, as explained in \cref{ssec:ispnetwork}.}
\vspace{-0.3cm}
\label{fig:globallocal}
\end{figure*}

The forward ISP network consists of four subnetworks: CanoNet, LocalNet, GlobalNet, and ParamNet (\cref{fig:overview}(a)).
CanoNet performs canonical ISP operations: demosaicing~\cite{getreuer2011malvar}, white balance, and color space conversion, exploiting the canonical parameters.
LocalNet performs local operations of an ISP such as denoising, sharpening and local tone mapping in addition to compensating for the residual errors of the canonical ISP operations of CanoNet.
GlobalNet performs global tone manipulation with respect to the camera parameters.
ParamNet is connected to LocalNet and GlobalNet and controls their behavior.
In the following, we explain each of CanoNet, LocalNet and GlobalNet in detail.

\paragraph{CanoNet}
CanoNet takes a RAW image $I_{RAW}\!\in\!\mathbb{R}^{1\times H\times W}$ and first performs demosaicing~\cite{getreuer2011malvar} to produce $I_{Dem}\!\in\!\mathbb{R}^{3\times H\times W}$ according to the Bayer pattern of $I_{RAW}$.
The demosaicing algorithm of CanoNet may differ from that of the target camera ISP.
Nonetheless, such discrepancy is compensated by the subsequent LocalNet.
It then performs white balance using the coefficients $g_{\mathrm{WB}}\in\mathbb{R}^{3}$.
Lastly, it transforms the color space from the RAW space to the linear sRGB space using the color correction matrix $M_{\mathrm{Cam}}\!\in\!\mathbb{R}^{3\times 3}$.
We denote the resulting image as $I_{lin}\!\in\!\mathbb{R}^{3\times H\times W}$.

\paragraph{LocalNet}
\cref{fig:globallocal} shows an overview of LocalNet.
LocalNet takes $I_{lin}$ and an optical parameter feature vector $z$ from ParamNet as input, and performs local ISP operations.
As well as $I_{lin}$, LocalNet also takes additional handcrafted features: a gradient map, a soft histogram map, and an over-exposure mask computed from $I_{lin}$, as they help improve reconstruction quality, as shown in \cite{lin2004radiometric,lin2005determining,liu2020single}.
For more details, refer to the supplementary material.
Finally, LocalNet predicts a residual image, which is then added to $I_{lin}$ to obtain the output image $I_{local}$.

As the goal of LocalNet is to learn local ISP operations, LocalNet adopts a network architecture based on the U-Net~\cite{ronneberger2015u}, which has been proven to be highly effective for various image restoration and enhancement tasks~\cite{zamir2021multi,wang2022uformer,abuolaim2020defocus,kupyn2018deblurgan,Chen2018DeepPhotoEnhancer}.
Specifically, the input image and the additional input features computed from the input image are concatenated, and fed to LocalNet. Besides, the optical parameter feature vector $z$ is duplicated horizontally and vertically to build a feature map of the same spatial size as the input image. The feature map is then added to an intermediate feature map of LocalNet.
Then, the resulting feature map is processed through multi-scale residual blocks (ResBlocks) with convolutional block attention modules (CBAMs)~\cite{woo2018cbam}, and converted to a residual image.

\paragraph{GlobalNet}
GlobalNet globally adjusts tone and color according to the content of its input image and the optical parameter feature vector $z$.
To this end, GlobalNet adopts parametric global operations and predicts their parameters.
Specifically, GlobalNet adopts two types of parametric global operations: quadratic transformation $f_q$ and gamma correction $f_g$.
Let us denote the input image to global adjustment operation as $I$, and its pixel as $p=(p_r,p_g,p_b)$.
Then, the quadratic transformation $f_q$ is defined as:
\begin{equation}
   f_q(p) = W p'
\end{equation}
where $p'$ is a 10-dim.~vector defined as $p'=[p_r^2,p_g^2,p_b^2,p_r p_g,p_g p_b,p_b p_r,p_r,p_g,p_b,1]^T$. $W$ is a $3\times10$ matrix with the coefficients of the quadratic transformation, which is uniformly applied to all the pixels in $I$.
The gamma correction operator $f_g$ is defined as:
\begin{equation}
    f_{g,c}(p_c) = \frac{(\alpha_c p_c + \beta_c)^{\gamma_c} - \beta_c^{\gamma_c}}{(\alpha_c + \beta_c)^{\gamma_c} - \beta_c^{\gamma_c}}
\end{equation}
where the subscript $c$ is an index to each color channel, i.e., $c\in\{r,g,b\}$.
$\gamma_c$ is a gamma parameter, while $\alpha_c$ and $\beta_c$ are a scale and an offset, respectively.
The set of the gamma correction coefficients is denoted by $G$, i.e., $G=\{\alpha_r,\beta_r,\gamma_r,\alpha_g,\beta_g,\gamma_g,\alpha_b,\beta_b,\gamma_b\}$.
Similar to $W$, $G$ is also uniformly applied to all the pixels in $I$.
For the sake of notational simplicity, we represent the quadratic transformation and gamma correction of an entire image $I$ as $f_q(I)$ and $f_g(I)$, respectively, in the rest of the paper.

To support a wide range of potential non-linear tone adjustments of commodity camera ISPs,
GlobalNet models the global tone adjustment as a series of gamma correction and quadratic transformation. 
Specifically, given an input image $I$, GlobalNet performs global tone adjustment as:
\begin{equation}
    \hat{I}=f_{gq}^N \circ\cdots\circ f_{gq}^1(I)
\end{equation}
where $f_{gq}^n(I)$ is defined as $f_{gq}^n=f_g^n(f_q^n(I))$.
$f_g^n$ and $f_q^n$ are the gamma correction and quadratic transformation with the $n$-th coefficients $G^n$ and $W^n$, respectively.
In our experiments, we use $N=4$ as it leads to the best quality (Refer to Tab.\capred{~S3} in the supplementary material).

The quadratic transformation is also adopted by previous modular learnable ISPs~\cite{afifi2021cie,schwartz2018deepisp}.
However, a single quadratic transformation cannot accurately model diverse non-linear tone adjustments performed by commodity camera ISPs including gamma correction.
Therefore, in our approach, we extend the global tone adjustment by adopting the gamma correction and iteratively applying both gamma correction and quadratic transformation.

\cref{fig:globallocal} shows an overview of GlobalNet.
To predict the coefficients $\{\cdots, (G^n,W^n), \cdots\}$, GlobalNet takes a concatenation of $I_{local}$ and the handcrafted features computed from $I_{local}$ as done in LocalNet as input, and extracts a global feature vector through a series of convolution layers with max pooling, and a global average pooling layer.
The global feature vector is then added to the optical parameter feature vector, and processed through fully-connected layers to obtain the coefficients.

\subsection{Inverse ISP Network}
The inverse ISP network also consists of CanoNet, LocalNet, GlobalNet and ParamNet with an inverse order of the forward ISP network (\cref{fig:overview}(b)). Specifically, GlobalNet takes an sRGB image as input and performs inverse tone manipulation. Next, LocalNet executes inverse local operations, and finally, CanoNet handles inverse color space conversion, inverse white balance, and mosaicing. GlobalNet and LocalNet use the same network architectures as those in the forward ISP network. On the other hand, CanoNet consists of the inverse operations of those of the forward ISP network, but in reverse order.

\section{Experiments}
\label{sec:exp}

\paragraph{Implementation}
We used PyTorch~\cite{paszke2017automatic} to implement our models. 
We train our model in two stages: pre-training and fine-tuning.
In the pre-training stage, we train our models with multiple datasets captured from multiple cameras.
In the fine-tuning stage, we train our models on a specific target camera.
Although our models aim at learning the ISP of a single target camera model, we found that pre-training with multiple cameras substantially improves the reconstruction quality.
In the pre-training stage, we use the RAISE dataset~\cite{dang2015raise} from Nikon D7000, D90, and D40, the RealBlur dataset~\cite{rim2020real} from Sony A7R3, and the S7 ISP dataset~\cite{schwartz2018deepisp} from Samsung Galaxy S7.
More details such as the effect of the pre-training, the statistics of the datasets, and how we split the datasets to training, validation, and test sets can be found in the supplementary material.

\begin{table}[t!]
    \centering
    \scalebox{0.8}{
    \begin{tabularx}{1.17\columnwidth}{c|c|c|cc}
    \Xhline{2.5\arrayrulewidth}
    \multicolumn{3}{c|}{Baseline (Ours w/o ParamNet)} & \multirow{3}{*}{PSNR$\uparrow$} & \multirow{3}{*}{Param$\downarrow$} \\ \cline{1-3}
    ParamNet & Non-linear & Random &  & \\ 
    (w/ Opt. Params) & Equalization & Dropout &  &  \\ \hline
                       &                &                      & 34.77                 & 0.68M                       \\ 
    $\checkmark$       &                &                      & -                     & -                           \\
    $\checkmark$       & $\checkmark$   &                      & \snd{35.64}                 & 0.71M                       \\
    $\checkmark$       & $\checkmark$   & $\checkmark$         & \fst{36.21}        & 0.71M                       \\
    \Xhline{2.5\arrayrulewidth}
    \end{tabularx}
    }
    \vspace{-0.2cm}
    \caption{Ablation study on the components of ParamNet.}
    \vspace{-0.1cm}
    \label{tab:abl-paramnet}
\end{table}

\begin{table}[t]
    \centering
    \scalebox{0.8}{
    \begin{tabularx}{1.12\columnwidth}{cccccc}
    \Xhline{2.5\arrayrulewidth}
    Otical Params & w/o $A$ & w/o $B$ & w/o $C$ & w/o $D$ & Full \\ \hline
    PSNR$\uparrow$ & 35.13 & 35.25 & \snd{35.87} & 35.46 & \fst{36.21} \\  
    SSIM$\uparrow$ & 0.9678 & 0.9718 & \snd{0.9724} & 0.9705 & \fst{0.9724} \\           
    \Xhline{2.5\arrayrulewidth}
    \end{tabularx}
    }
    \vspace{-0.2cm}
    \caption{Ablation study on the impact of each optical parameter. $A$,$B$,$C$, and $D$ represent sensor sensitivity, exposure time, aperture size, and focal length, respectively.}
    \vspace{-0.4cm}
    \label{tab:abl-optparams}
\end{table}

We train our network in the pre-training stage for 520 epochs with an initial learning rate of $2.0\times10^{-4}$, and in the fine-tuning stage for 2,140 epochs with an initial learning rate of $2.0\times10^{-5}$.
For both pre-training and fine-tuning, we employ AdamW~\cite{loshchilov2017decoupled} and the learning rate is step-decayed with a rate of 0.8 every 10 epochs.
In each epoch in the pre-training stage, we randomly sample 1024 cropped patches of size $448\times448$ from the dataset of each camera, resulting in a total of 5120 patches.
In the fine-tuning stage, we sample 1024 patches from the dataset of a target camera in each epoch.
We evaluated our models and other models on a PC equipped with an NVIDA RTX A6000.
For a fair comparison, we also apply both pre-training and fine-tuning to other models in our experiments.

We train our forward and inverse ISP networks in an end-to-end fashion using losses $\mathcal{L}_{for}$ and $\mathcal{L}_{inv}$:
\begin{eqnarray}
    \mathcal{L}_{for} &=& \|\hat{I}_{sRGB} - I_{sRGB}\|_1 \\
    \mathcal{L}_{inv} &=& \|\hat{I}_{RAW} - I_{RAW}\|_1
\end{eqnarray}
respectively,
where $\hat{I}_{sRGB}$, $I_{sRGB}$, $\hat{I}_{RAW}$, and $I_{RAW}$ are a reconstructed sRGB image, its ground-truth sRGB image, a reconstructed RAW image, and its ground-truth RAW image, respectively.

\subsection{Ablation Study}

\begin{figure*}[t!]
\includegraphics[width=1\linewidth]{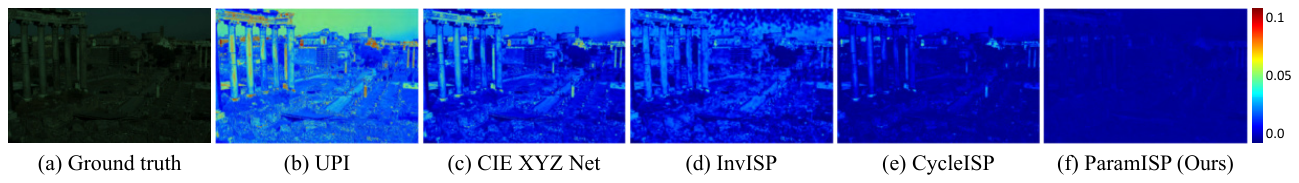}
\vspace{-0.6cm}
\caption{sRGB-to-RAW reconstruction. We show error maps between reconstructed and GT RAW images.}
\vspace{-0.2cm}
\label{fig:inverse}
\end{figure*}

\begin{figure*}[t]
\includegraphics[width=1\linewidth]{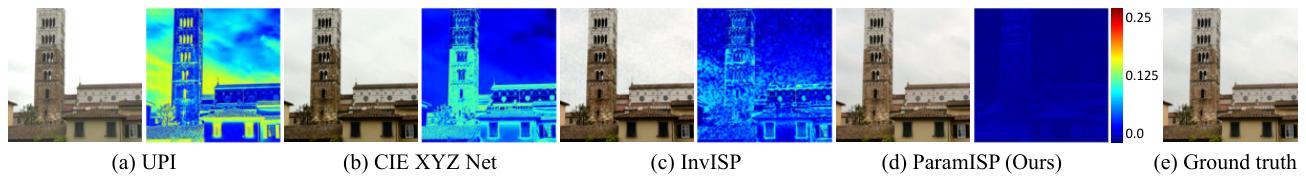}
\vspace{-0.6cm}
\caption{RAW-to-sRGB reconstruction. We show error maps between reconstructed and GT sRGB images.}
\vspace{-0.3cm}
\label{fig:forward}
\end{figure*}

For all the ablation studies in this section, we train inverse (sRGB-to-RAW) ISP network using the D7000 training images of the RAISE dataset~\cite{dang2015raise}.
Additional ablation studies using the forward ISP network and different camera datasets, and on the effect of the input features and training approach can be found in the supplementary material.

\paragraph{ParamNet}
We first evaluate the effect of the non-linear equalization and random optical parameter dropout of ParamNet.
To this end, we compare variants of our model in \cref{tab:abl-paramnet}.
The first row in the table corresponds to a baseline model without ParamNet.
Introducing ParamNet without non-linear equalization (2nd row) results in the failure of training with diverging losses.
On the other hand, the non-linear equalization (3rd row) not only stabilizes the training but also improves the reconstruction quality compared to the baseline model in the first row,
which proves that the non-linear equalization is essential for training, and that exploiting the optical parameters is necessary to improve the reconstruction quality.
Finally, random dropout of the optical parameters (4th row) further enhances the reconstruction quality as it prevents overfitting and enhances the generalization ability of ParamNet.

\paragraph{Optical Parameters}
We now analyze the impact of the optical parameters.
To this end, we compare four variants of ParamNet, for each of which, we exclude each one of the four optical parameters.
\cref{tab:abl-optparams} shows the quantitative ablation results.
Compared to the model that uses all the optical parameters, all the other variants exhibit performance degradation, indicating that all the optical parameters need to be considered for high-quality reconstruction.
We can also observe that excluding the sensitivity and exposure time shows the largest performance degradation.
While how commodity camera ISPs utilize the optical parameters is unknown, these optical parameters may affect the image enhancement operation of camera ISPs, such as denoising, as they are closely related to noise.
The impact of the aperture size and focal length is not as significant as the others, but the reconstruction quality still decreases without them.
This is because aperture size and focal length are related to defocus blur and lens aberrations, so the camera ISP operations may alter based on them.

\begin{table}[t!]
    \centering
    \scalebox{0.8}{
    \begin{tabularx}{1.08\columnwidth}{c|c|ccc}
    \Xhline{2.5\arrayrulewidth}
    \multicolumn{2}{c|}{Baseline (CIE XYZ Net~\cite{afifi2021cie})} & \multirow{2}{*}{PSNR$\uparrow$} & \multirow{2}{*}{SSIM$\uparrow$}  & \multirow{2}{*}{Param$\downarrow$} \\ \cline{1-2}
    \,\,\,\,LocalNet\,\,\,\, & GlobalNet & & & \\ \hline
                   &                      & 30.04                & 0.9461                 & 1.3M                       \\ 
    $\checkmark$   &                      & \snd{33.12}          & \snd{0.9644}           & 1.7M                       \\
    $\checkmark$   & $\checkmark$         & \fst{33.66}          & \fst{0.9646}           & 0.6M              \\
    \Xhline{2.5\arrayrulewidth}
    \end{tabularx}
    }
    \vspace{-0.2cm}
    \caption{Ablation study on the effects of LocalNet and GlobalNet.}
    \vspace{-0.4cm}
    \label{tab:abl-ispnetwork}
\end{table}

\paragraph{ISP Network Architecture}
We verify the effectiveness of the network architecture of our ISP subnetworks.
As our modular network structure is inspired by CIE XYZ Net~\cite{afifi2021cie} that consists of a pair of local and global mapping modules, we consider its local and global mapping networks as our baseline for LocalNet and GlobalNet.
Specifically, we build a baseline model for the inverse ISP network by replacing the LocalNet and GlobalNet with the local and global mapping networks of CIE XYZ Net, then compare the performance of the baseline and its variants to ours.
The local mapping network of CIE XYZ Net simply stacks convolution layers, while the global mapping network models global tone manipulation as a single quadratic function and estimate its parameters using fully-connected layers.
\cref{tab:abl-ispnetwork} shows the comparison result. 
The table shows that LocalNet significantly improves the PSNR by more than 3 dB over the local mapping of CIE XYZ Net.
Moreover, GlobalNet further improves the PSNR by around 0.5 dB and leads to a significantly reduced model size thanks to our global tone adjustment model and efficient network architecture.

\subsection{RAW \& sRGB Reconstruction}
\label{ssec:raw-recon}

In this section, we compare the sRGB-to-RAW and RAW-to-sRGB reconstruction ability of ParamISP with existing state-of-the-art methods: UPI~\cite{brooks2019unprocessing}, CIE XYZ Net~\cite{afifi2021cie}, CycleISP~\cite{zamir2020cycleisp}, and InvISP~\cite{xing2021invertible}.
UPI is a non-learnable parametric method that approximates the general structure of the camera ISP.
CIE XYZ Net, CycleISP, and InvISP are learning-based approaches that employ symmetrical forward and inverse ISP networks.
For CIE XYZ Net whose output is an image in the CIE XYZ color space, we convert its output to a RAW image using canonical ISP operations as in CanoNet for a fair comparison.
For InvISP whose output is white-balanced RAW image, we apply inverse white balancing and mosaicing operations to obtain a RAW image.
These learning-based methods train the forward and inverse networks together, but we found that separately training them for each yields better performance. Therefore, we train them separately in the same manner as ours.

\begin{table}[t!]
\aboverulesep = 0.11mm 
\belowrulesep = 0.1mm 
    \centering
    \scalebox{0.8}{
    \begin{tabularx}{1.25\columnwidth}{lcccccc}
    \Xhline{2.5\arrayrulewidth}
\multirow{2}{*}{Method}& \multicolumn{5}{c}{sRGB$\rightarrow$RAW measured in PSNR$\uparrow$} & \multirow{2}{*}{Param$\downarrow$} \\
    \cmidrule(lr){2-6}
    & D7000 & D90   & D40   & S7    & A7R3  & \\
    \midrule
    UPI~\cite{brooks2019unprocessing} & 20.67 & 26.57 & 22.05 & 29.98 & 30.48 & -    \\
    CIE XYZ Net~\cite{afifi2021cie}   & 30.04 & 32.62 & 38.57 & 33.24 & 36.42 & 1.3M \\
    CycleISP~\cite{zamir2020cycleisp} & \snd{35.52} & \snd{35.85} & 42.83 & \snd{34.55} & 45.35 & 3.1M \\
    InvISP~\cite{xing2021invertible}  & 33.48 & 35.39 & \snd{45.08} & 34.29 & \snd{47.14} & 1.4M \\
    ParamISP (Ours) & \fst{38.49} & \fst{37.06} & \fst{45.97} & \fst{35.20} & \fst{48.33} & 0.7M \\
    \Xhline{2.5\arrayrulewidth}
    \end{tabularx}
    }
    \vspace{-0.2cm}
    \caption{Quantitative comparison on RAW reconstruction.}
    \vspace{-0.4cm}
    \label{tab:rawrecon-results}
\end{table}

\cref{tab:rawrecon-results} shows quantitative results of the inverse sRGB-to-RAW reconstruction in terms of PSNR evaluated on our test set.
UPI~\cite{brooks2019unprocessing} results in high reconstruction errors due to its parametric model-based pipeline.
CIE XYZ Net~\cite{afifi2021cie}, CycleISP~\cite{zamir2020cycleisp}, and InvISP~\cite{xing2021invertible} show better reconstruction quality than UPI, but their quality is still limited compared to our method.
ParamISP achieves the best reconstruction quality despite its much smaller model size, thanks to leveraging the optical parameters and our carefully-designed network architecture.
\cref{fig:inverse} shows the qualitative results of error maps between reconstructed RAW images and ground-truth RAW images.
As the figure shows, our method produces much less error than the others.

\cref{tab:rgbrecon-results} reports quantitative results of the forward reconstruction (RAW-to-sRGB).
CycleISP is not included in this comparison because it needs an input sRGB image for sRGB reconstruction in the sRGB-to-RAW-to-sRGB cyclic reconstruction.
Similar to the inverse sRGB-to-RAW reconstruction, ParamISP clearly outperforms the other methods, showing the effectiveness of our approach.

\begin{table}[t!]
\aboverulesep = 0.11mm 
\belowrulesep = 0.1mm 
    \centering
    \scalebox{0.8}{
    \begin{tabularx}{1.25\columnwidth}{lcccccc}
    \Xhline{2.5\arrayrulewidth}
    \multirow{2}{*}{Method}& \multicolumn{5}{c}{RAW$\rightarrow$sRGB measured in PSNR$\uparrow$} & \multirow{2}{*}{Param$\downarrow$} \\
    \cmidrule(lr){2-6}
    & D7000 & D90   & D40   & S7    & A7R3  & \\
    \midrule
    UPI~\cite{brooks2019unprocessing} & 18.81 & 20.30 & 16.01 & 20.05 & 19.37 & -    \\
    CIE XYZ Net~\cite{afifi2021cie}   & 26.76 & 27.61 & 34.84 & 27.63 & 37.19 & 1.3M \\
    InvISP~\cite{xing2021invertible}  & \snd{30.20} & \snd{28.89} & \snd{37.86} & \snd{28.96} & \snd{43.93} & 1.4M \\
    ParamISP (Ours) & \fst{34.14} & \fst{30.83} & \fst{39.54} & \fst{29.02} & \fst{45.51} & 0.7M \\
    \Xhline{2.5\arrayrulewidth}
    \end{tabularx}
    }
    \vspace{-0.2cm}
    \caption{Quantitative comparison on sRGB reconstruction.}
    \vspace{-0.4cm}
    \label{tab:rgbrecon-results}
\end{table}

\section{Applications}
\label{sec:app}

Forward and inverse ISP models can benefit various applications as shown in \cite{afifi2021cie,xing2021invertible,zamir2020cycleisp}.
In this section, we demonstrate a couple of applications of ParamISP:  RAW deblurring and HDR reconstruction.
Other additional applications including deblurring dataset synthesis and camera-to-camera transfer are also provided in the supplementary material.
Before applying ParamISP to applications, we further perform joint fine-tuning on separately trained forward and inverse ISP networks.
For details on the joint fine-tuning, refer to the supplementary material.

\begin{figure}[t!]
\includegraphics[width=1.0\linewidth]{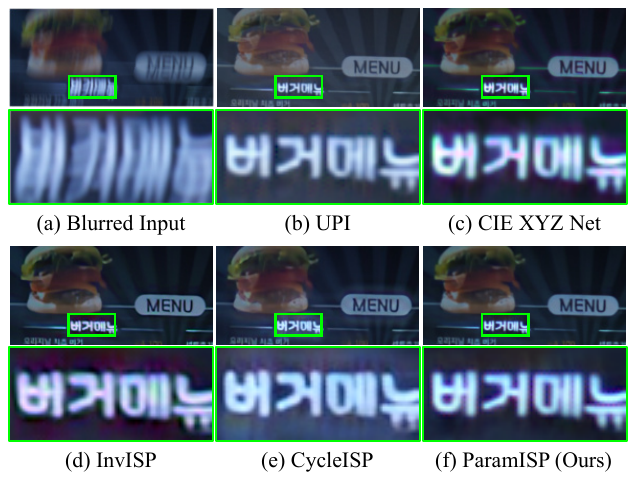}
\centering
\vspace{-0.6cm}
\caption{Qualitative results of RAW deblurring. In (c) and (d), red artifacts are visible in white text, while in (b) and (e), the results are blurrier than ours.}
\vspace{-0.2cm}
\label{fig:rawdeblur}
\end{figure}

\paragraph{Deblurring in RAW Space}
While the irradiance of a blurred image has a linear relationship with its latent sharp image, such a linear relationship no longer holds in sRGB images after non-linear operations have been applied by camera ISPs. Such non-linearity varies across different cameras.
Consequently, learning sRGB image deblurring requires camera-specific training datasets, which leads to an excessive amount of training time and memory space to support diverse camera models.
Instead, we can effectively reconstruct a RAW image from an sRGB image using ParamISP, apply a camera-independent deblurring model to the RAW image, and obtain a deblurred sRGB image.
While ParamISP also requires camera-wise training, it requires a much smaller number of parameters ($2\times0.7$M) compared to deblurring models (Stripformer: 20M~\cite{Tsai2022Stripformer} and Uformer-B: 51M~\cite{wang2022uformer}), and supports a wide range of applications.

\cref{tab:rawdeblur} and \cref{fig:rawdeblur} show quantitative and qualitative comparisons of deblurring results in the RAW space using different ISP networks.
For deblurring in the RAW space, we use Stripformer~\cite{Tsai2022Stripformer} trained on the RealBlur-R dataset, which is a real-world RAW blurry image dataset~\cite{rim2020real}.
The evaluation is done on the RealBlur-J test set~\cite{rim2020real}.
As the comparisons show, ParamISP clearly outperforms all the other ISP networks in deblurring performance thanks to its high-quality sRGB and RAW reconstruction.

\begin{table}[t]
    \centering
    \scalebox{0.8}{
    \begin{tabularx}{1.23\columnwidth}{cccccc}
    \Xhline{2.5\arrayrulewidth}
    \multirow{2}{*}{Method} & \multirow{2}{*}{\shortstack{UPI\\~\cite{brooks2019unprocessing}}} & \multirow{2}{*}{\shortstack{CIE XYZ Net\\~\cite{afifi2021cie}}} & \multirow{2}{*}{\shortstack{InvISP\\~\cite{xing2021invertible}}} & \multirow{2}{*}{\shortstack{CycleISP\vspace{-0.07cm}\\~\cite{zamir2020cycleisp}}} & \multirow{2}{*}{\shortstack{ParamISP\\(Ours)}} \\  
     & & & & & \\ \hline
    PSNR$\uparrow$    & 24.63  & 28.30  & 28.82  & \snd{29.94}  & \fst{30.70}  \\ 
    SSIM$\uparrow$    & 0.8011 & 0.8444 & 0.8721 & \snd{0.8855} & \fst{0.8984} \\
    Param$\downarrow$   & -      & 2.7M   & 1.4M   & 7.4M   & 1.4M   \\
    \Xhline{2.5\arrayrulewidth}
    \end{tabularx}
    }
    \vspace{-0.2cm}
    \caption{Quantitative results of RAW deblurring.}
    \vspace{-0.4cm}
    \label{tab:rawdeblur}
\end{table}

\paragraph{HDR Reconstruction}
As shown in CIE XYZ Net~\cite{afifi2021cie}, RAW image reconstruction can also be used for high-dynamic-range (HDR) image reconstruction from a single low-dynamic-range (LDR) image as RAW images provide a wider tonal value range.
Specifically, we first convert an LDR sRGB image to an LDR RAW image, and multiply the reconstructed RAW image by synthetic digital gains (0.1, 1.4, 2.7, 4.0) to create four RAW images.
Each of these is then passed through the forward ISP to reconstruct sRGB images. Finally, we apply an off-the-shelf exposure-fusion algorithm~\cite{mertens2007exposure} to obtain a single HDR image.

\begin{figure}[t!]
\includegraphics[width=1.0\linewidth]{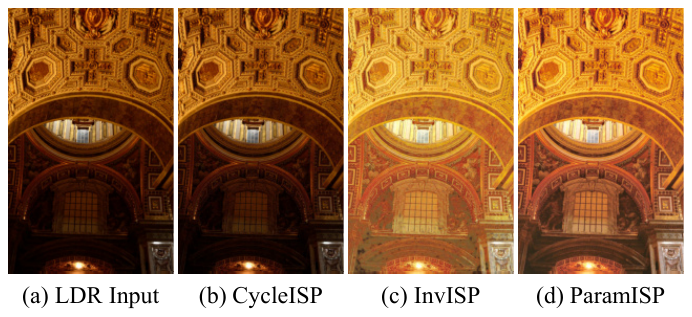}
\centering
\vspace{-0.6cm}
\caption{Qualitative results of HDR reconstruction.}
\vspace{-0.4cm}
\label{fig:HDR}
\end{figure}

\cref{fig:HDR} shows qualitative results using different models.
Previous methods primarily focus on cyclic reconstruction.
Specifically, CycleISP uses an input LDR sRGB image for the forward ISP to adjust the tone, while InvISP, composed of a single flow-based network, is not robust when intermediate RAW images are altered.
On the other hand, ParamISP successfully produces an HDR reconstruction result compared to other methods.
For more qualitative results, refer to the supplementary material.

\section{Conclusion}
In this paper, we present ParamISP, a novel learning-based forward and inverse ISP framework that leverages camera parameters.
To incorporate camera parameters effectively, we introduce ParamNet to control the forward and inverse ISP networks, proposing a stable and effective training strategy with respect to camera parameters.
We also present novel network architectures for ISP networks that better reflect real-world ISP operations.
Through our extensive experiments, ParamISP shows the state-of-the-art performance in RAW and sRGB reconstruction 
and the robust applicability to various applications.

\paragraph{Limitations and Future Work}
While we have quantitatively confirmed that incorporating the aperture size and focal length improves the reconstruction quality in \cref{tab:abl-optparams}, it is still unclear how such parameters affect the behaviors of the ISP operations.
While we use datasets captured by various cameras in the pre-training stage, different ISPs may behave differently with respect to different optical parameters, and such inconsistency may potentially have negative impact on a pre-trained ISP model.
A more sophisticated training strategy may resolve such issue.

\paragraph{Acknowledgments}
This work was supported by the NRF grants (RS-2023-00211658, RS-2023-00280400, 2022R1A6A1A03052954, 2023R1A2C200494611) and IITP grant (2019-0-01906, Artificial Intelligence Graduate School Program (POSTECH)) funded by the Korea government (MSIT). This work was also supported by Samsung Research Funding Center (SRFCIT1801-52) and Samsung Electronics Co.

{
    \small
    \bibliographystyle{ieeenat_fullname}
    \bibliography{references}
}


\end{document}

% make title and teaser
\twocolumn[{
\renewcommand\twocolumn[1][]{#1}
\maketitle
\begin{center}
    \centering
    \captionsetup{type=figure}
    \includegraphics[width=0.98\textwidth]{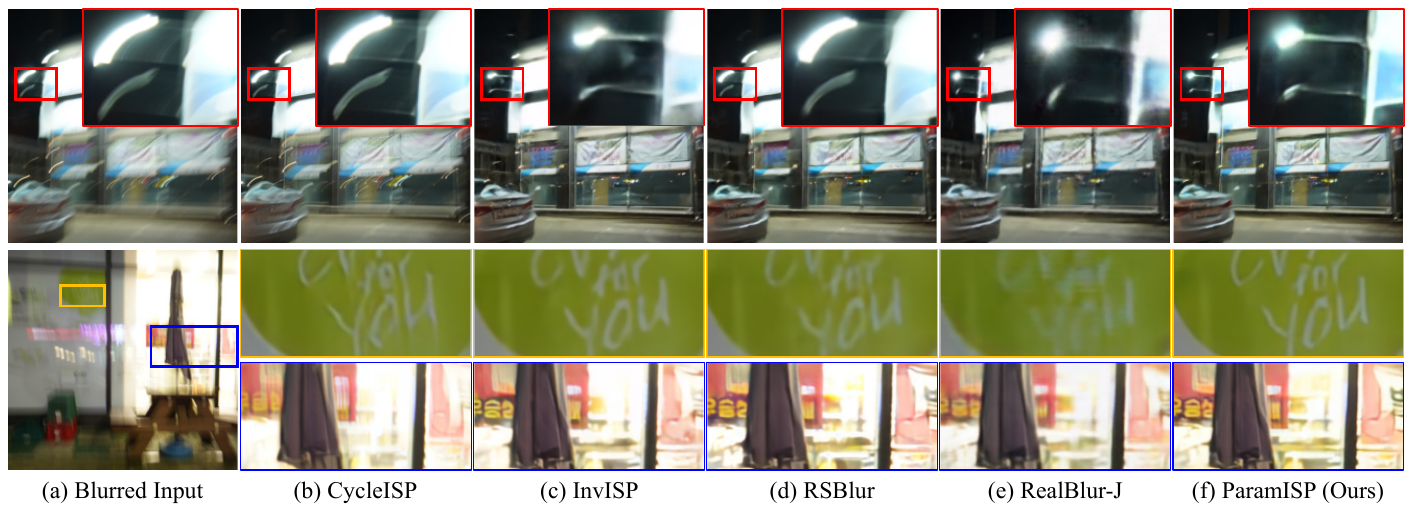}
    \captionof{figure}{Qualitative results of deblurring dataset synthesis. 
    Compared to other methods, the deblurring model~\cite{nafnet} trained on the dataset synthesized by our ISP models more effectively restores sharp details.}
    \label{fig:deblurdataset}
\end{center}
}]

\blfootnote{$^{\ast}$ Equal contribution.}
\blfootnote{$^{\dagger}$ Work done prior to joining Samsung.}
\section{Overview}
In this supplementary material, we describe two additional applications: deblurring dataset synthesis and camera-to-camera transfer (\cref{sec:applications-supp}). 
We then provide details and an ablation study on the input features used in LocalNet and GlobalNet (\cref{sec:infeatures-supp}), as well as additional experiments on the global adjustment operation in GlobalNet (\cref{sec:globalnet-supp}) and additional discussions on the training strategy (\cref{sec:strategy-supp}).
We also provide experimental results on the relationship between optical parameters and ParamNet (\cref{sec:opticalparams-supp}).
Finally, we present additional quantitative and qualitative results (\cref{sec:moreresults-supp}), along with the detailed architecture of ParamISP (\cref{sec:networkarch-supp}).

\section{Additional Applications}
\label{sec:applications-supp}
ParamISP can be applied to various applications (\eg, deblurring dataset synthesis, RAW deblurring, HDR reconstruction, and camera-to-camera transfer) unlike previous methods.
In this section, we describe two additional applications (\ie, deblurring dataset synthesis, camera-to-camera transfer) that were not covered in the main paper.
We find that joint fine-tuning on separately trained forward and inverse ISP networks brings additional performance improvements in applications. 
Therefore, before applying ParamISP to applications excluding camera-to-camera transfer, we conduct additional joint fine-tuning.

In the joint fine-tuning stage, we train our pretrained forward and inverse ISP networks for 450 epochs with an initial learning rate of $1.0\times10^{-4}$.
We jointly fine-tune separately trained forward and inverse ISP networks in an end-to-end manner using loss $\mathcal{L}_{joint}$:
\begin{align}
    \mathcal{L}_{joint} & = \|f_{for}(f_{inv}(I_{sRGB})) - I_{sRGB}\|_1 \\ 
                        & \,\,\,\,\,\,\,\,\,\, + \|f_{inv}(I_{sRGB}) - I_{RAW}\|_1, \nonumber
\end{align}
where $f_{for}$ and $f_{inv}$ are the forward and inverse ISP networks, respectively, and $I_{sRGB}$ is an sRGB image.
Other training conditions are the same as those explained in Sec.~4~of the main paper.

\subsection{Deblurring Dataset Synthesis}
\begin{table}[t!]
    \centering
    \scalebox{0.8}{
    \begin{tabularx}{1.205\columnwidth}{c|cccc|c}
    \Xhline{2.5\arrayrulewidth}
    \multirow{3}{*}{Method} & \multicolumn{4}{c|}{Synthetic} & Real \\ \cline{2-6}
     & \multirow{2}{*}{\shortstack{CycleISP \vspace{-0.07cm} \\ ~\cite{zamir2020cycleisp}}} & \multirow{2}{*}{\shortstack{InvISP\\~\cite{xing2021invertible}}} & \multirow{2}{*}{\shortstack{RSBlur\\~\cite{rim_2022_ECCV}}} & \multirow{2}{*}{\shortstack{ParamISP\\(Ours)}} & \multirow{2}{*}{\shortstack{RealBlur-J\\~\cite{rim2020real}}} \\  
     & & & & & \\ \hline
    PSNR$\uparrow$    & 29.98  & 31.06  & 31.16  & \snd{31.31}  & \fst{31.82}  \\ 
    SSIM$\uparrow$    & 0.8806 & 0.9134 & 0.9142 & \snd{0.9148} & \fst{0.9203} \\
    \Xhline{2.5\arrayrulewidth}
    \end{tabularx}
    }
    \vspace{-0.2cm}
    \caption{Quantitative results of deblurring dataset synthesis. Our dataset synthesis outperforms all the other synthesis methods.
    While ours still achieves lower performance than the RealBlur-J training set, this is partly due to the photometric misalignment present in the RealBlur-J dataset.}
    \vspace{-0.4cm}
    \label{tab:deblurdataset}
\end{table}

It is essential to reflect the camera ISP in order to synthesize a realistic image restoration dataset such as deblurring datasets~\cite{rim_2022_ECCV}.
ParamISP can enhance the accuracy of synthetic deblurring datasets as it can more accurately model real-world camera ISPs.
\cref{tab:deblurdataset} shows a comparison among different dataset synthesis approaches.
On the table, RSBlur~\cite{rim_2022_ECCV} is a baseline model that uses a simple parametric ISP model, while InvISP~\cite{xing2021invertible}, CycleISP~\cite{zamir2020cycleisp}, and ParamISP mean variants of the RSBlur pipeline whose ISP model is replaced by the corresponding ISP models.
For fair comparisons, we train the other ISP models according to their respective learning approaches.
We synthesize deblurring datasets using each of the approaches on the table, train a deblurring model~\cite{nafnet} using each dataset, and evaluate their performance on the RealBlur-J test set~\cite{rim2020real}, which is a real-world blur dataset.
The last column of the table represents the results obtained by directly training the deblurring model on the RealBlur-J train set, and we consider this as the upper bound.

As \cref{tab:deblurdataset} shows, ParamISP achieves the closest PSNR to the upper bound, outperforming all the other synthesis approaches.
Interestingly, InvISP~\cite{xing2021invertible} and CycleISP~\cite{zamir2020cycleisp} achieve worse performance than RSBlur~\cite{rim_2022_ECCV} although they are learnable approaches.
This is because they primarily focus on cyclic reconstruction (sRGB-to-RAW-to-sRGB) and are less suitable for applications that manipulate RAW images as described in the main paper.
Specifically, CycleISP uses the input sRGB image for restoring the tone when reconstructing an sRGB image back from a RAW image.
Here, for the smooth tone restoration of CycleISP, we apply the same blur kernel used to create the blurry RAW image to the input sharp sRGB image, rather than using the input sharp sRGB image directly.
InvISP uses a single normalizing flow-based invertible network, resulting in near-perfect reconstruction quality for cyclic reconstruction. 
However, its quality significantly degrades when the intermediate RAW images are altered.

While our method outperforms all the other synthetic dataset generation processes both in PSNR and SSIM, it still achieves lower performance than the upper bound model trained using the RealBlur-J training set. We emphasize that the performance gap between the upper bound and ours is partly due to the existence of remaining photometric misalignment in the blurry and sharp image pairs in the RealBlur-J dataset. The sharp and blurry images of the RealBlur-J dataset were captured by different cameras, so they have slightly different tones. While the postprocessing process of the RealBlur-J dataset applies photometric alignment to mitigate this issue, the images in the dataset still have remaining tone difference, which could only be learned from the RealBlur-J training set.
Dataset synthesis methods that synthesize blurry images using only sharp images are unable to depict such tone difference between different cameras and, in fact, there is no need to depict such tone differences for the purpose of deblurring.
\cref{fig:deblurdataset} shows a qualitative comparison.
The deblurring model trained with ParamISP visually outperforms each deblurring model trained with other methods, including the upper bound RealBlur-J.

\subsection{Camera-to-Camera Transfer} 
\begin{figure}[t!]
\includegraphics[width=1.0\linewidth]{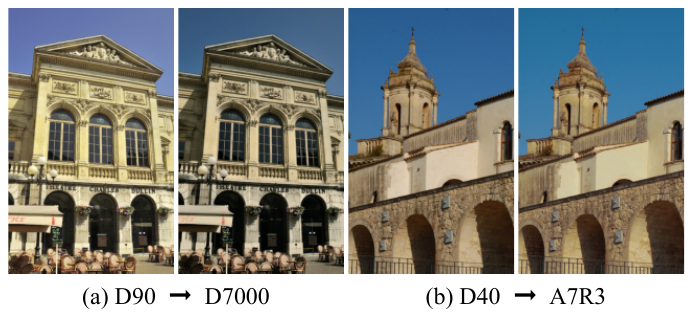}
\centering
\vspace{-0.6cm}
\caption{Qualitative results of camera-to-camera transfer.}
\vspace{-0.4cm}
\label{fig:cam2cam}
\end{figure}

Given an sRGB image, ParamISP can manipulate the image as if it was taken by a different camera using the inverse and forward ISP networks trained on the different camera.
\cref{fig:cam2cam} qualitatively shows the camera-to-camera transfer results of ParamISP. The resulting camera-transferred images (2nd and 4th images) show color tone changes from the input images (1st and 3rd images) without any noticeable artifacts.

\section{Input Features for LocalNet and GlobalNet}
\label{sec:infeatures-supp}

\begin{figure}[t]
\includegraphics[width=1.0\linewidth]{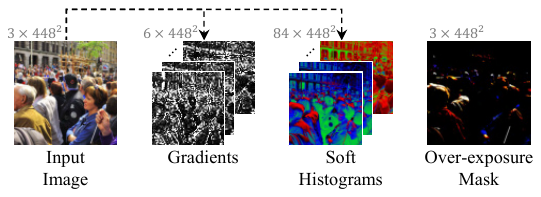}
\centering
\vspace{-0.6cm}
\caption{Input features of LocalNet and GlobalNet.}
\label{fig:infeatures}
\end{figure}

It is reported that leveraging the gradient map and color histogram maps improves the ISP network performance by several previous works~\cite{lin2004radiometric, lin2005determining,liu2020single}.
In our preliminary experiments, we also find a similar finding. Based on this, we design our LocalNet and GlobalNet to be fed additional input features as well as an input image.
As additional input features, we include a gradient map, a soft histogram map, and an over-exposure mask to improve the ISP network performance (\cref{fig:infeatures}).
For the gradient map, we apply the Sobel filter~\cite{kanopoulos1988design} on an input image and compute per-channel gradient maps in vertical and horizontal directions, resulting in a 6-channel gradient map.
For the soft histogram map, we compute the soft histogram~\cite{liu2020single}, for which we measure the relative distance between each channel value of a pixel and the center of histogram bins. In practice, we use 28 histogram bins, resulting in an 84-channel soft histogram map.
We also include an over-exposure mask as a hint for restoring pixels of range-clipped values, where we compute the mask by setting its values as $10\max(x-\tau,0)$ where $x$ is a pixel value of an input image, and $\tau$ is a threshold.
We use 0.9 as the threshold in our implementation.

\parhead{Ablation Study}
To validate the effects of the input features, we conduct an ablation study on RAW reconstruction using the D7000 images of the RAISE dataset~\cite{dang2015raise}.
\cref{tab:abl-infeatures} shows a quantitative ablation study on the input features of LocalNet and GlobalNet.
To analyze only the effects of the input features, we use our model without ParamNet as the baseline (4th column).
We prepare our baseline with its three model variants, where LocalNet and GlobalNet in each model variant do not use each one of the three input features.

It may be unnecessary to explicitly provide hand-crafted features such as image gradients and soft histograms, as a network can learn to extract such features from an input image.
However, without these features, the network may not fully exploit its capacity to learn features more useful for local and global non-linear operations, resulting in low reconstruction quality (1st and 2nd columns).
Furthermore, discarding an over-exposure mask may waste the network capacity in estimating and restoring pixels of range-clipped values, resulting in decreased reconstruction performance (3rd column).
Additionally, in the first and second rows of \cref{tab:abl-components}, we describe the results of experiments conducted with various cameras and forward ISP networks to assess the impact of input features.

\begin{table}[t]
    \centering
    \scalebox{0.8}{
    \begin{tabularx}{0.96\columnwidth}{ccccc}
    \Xhline{2.5\arrayrulewidth}
    Input Features & w/o $A$ & w/o $B$ & w/o $C$ & Full \\ \hline
    PSNR$\uparrow$ & 34.32 & 34.40 & 34.29 & \fst{34.77} \\  
    SSIM$\uparrow$ & 0.9693 & 0.9665 & 0.9677 & \fst{0.9712} \\           
    \Xhline{2.5\arrayrulewidth}
    \end{tabularx}
    }
    \vspace{-0.2cm}
    \caption{Ablation study on the impact of each input feature. $A$, $B$, and $C$ represent gradient map, soft histogram, and over-exposure mask, respectively.}
    \vspace{-0.4cm}
    \label{tab:abl-infeatures}
\end{table}

\section{Global Adjustment Operation of GlobalNet}
\label{sec:globalnet-supp}

\begin{table}[t]
    \centering
    \scalebox{0.8}{
    \begin{tabularx}{1.25\columnwidth}{cccccccc}
    \Xhline{2.5\arrayrulewidth}
    Operation & $A$ & $B$ & $C$$\times$1 & $C$$\times$2 & $C$$\times$3 & $C$$\times$4 & $C$$\times$5 \\ \hline
    PSNR$\uparrow$ & 32.70 & 32.77 & 32.87 & 32.96 & 33.27 & \fst{33.66} & \snd{33.50} \\  
    SSIM$\uparrow$ & 0.962 & 0.961 & 0.961 & 0.962 & 0.963 & \snd{0.965} & \fst{0.967} \\           
    \Xhline{2.5\arrayrulewidth}
    \end{tabularx}
    }
    \vspace{-0.2cm}
    \caption{Ablation study on the effects of the global adjustment operation of GlobalNet. $A$ and $B$ represent the 3$\times$6 polynomial mapping function of CIE XYZ Net~\cite{afifi2021cie} and the 3$\times$10 quadratic transformation of DeepISP~\cite{schwartz2018deepisp}, respectively, while $C\times N$ denotes $N$ pairs of gamma correction and quadratic transformation (Ours).}
    \label{tab:abl-funcnum}
\end{table}

In this section, we verify the effect of the global adjustment operations of GlobalNet.
To this end, we prepare variants of ParamISP with different global adjustment operations, and train them using the D7000 training images of the RAISE dataset~\cite{dang2015raise}. Then, we evaluate their performances using the D7000 test set (\cref{tab:abl-funcnum}).
In this evaluation, we include the global adjustment operations of previous methods~\cite{afifi2021cie,schwartz2018deepisp} as well as different numbers of gamma correction and quadratic transformation operations.
\cref{tab:abl-funcnum} shows that our global tone adjustment operation substantially improves the performance in PSNR and SSIM.

\begin{table}[t!]
    \centering
    \scalebox{0.8}{
    \begin{tabularx}{1.24\columnwidth}{cccccc}
    \Xhline{2.5\arrayrulewidth}
    Camera model & A7R3\cite{rim2020real} &D7000\cite{dang2015raise} & D90\cite{dang2015raise} & D40\cite{dang2015raise} & S7\cite{schwartz2018deepisp} \\ \hline
    Training \#    & 7766 & 4600 & 1700  & 26  & 50  \\ 
    Validation \#  & 200  & 200  & 100   & -   & 20  \\
    Testing \#     & 1000 & 1000 & 400   & 50  & 150 \\
    \Xhline{2.5\arrayrulewidth}
    \end{tabularx}
    }
    \vspace{-0.2cm}
    \caption{Statistics of our dataset used for ParamISP.}
    \vspace{-0.4cm}
    \label{tab:data-config}
\end{table}

\section{Training Strategy}
\label{sec:strategy-supp}

\parhead{Datasets}
we use three datasets consisting of RAW and sRGB image pairs captured from multiple cameras: the RAISE dataset~\cite{dang2015raise} from Nikon D7000, D90, and D40, the RealBlur dataset~\cite{rim2020real} from Sony A7R3; and the S7 ISP dataset~\cite{schwartz2018deepisp} from Samsung Galaxy S7.
The statistics of each dataset are shown in \cref{tab:data-config}.
We extract camera parameters from the EXIF metadata~\cite{tachibanaya2001description} included in JPEG images. 

\parhead{Pre-training with Images from Diverse Cameras}
To achieve high-quality reconstruction, we pre-train our models using datasets of multiple cameras as if they were captured by a single camera model.
We then fine-tune the models for our target camera.
Despite the differences across different camera models, we find that this two-stage training substantially improves the ISP performance as the ISP models can learn common knowledge on the ISP operations.

To validate our two-stage training approach, we present a quantitative ablation study result in \cref{tab:abl-strategy}.
In the table, `Generic', `Individual', and `Generic$+$individual' mean models trained using multiple camera datasets, models trained using only target camera datasets, and models trained using our two-stage training scheme, respectively.
All the `Generic', `Individual' and `Generic$+$individual' models are without ParamNet.
We also include our final model `ParamISP' trained using our two-stage training scheme in the table.

In the table, the `Individual' models show higher RAW and sRGB reconstruction performance than the `Generic' models on average as the `Generic' models cannot properly learn the behaviors of specific camera models.
The table also shows that the `Generic$+$individual' models achieve higher performance than both `Generic' and `individual' models as they can exploit common knowledge on the camera ISP operations across various camera models, and at the same time, they can accurately learn the behaviors of specific camera models.
Finally, our full models (ParamISP) outperform all the other models thanks to ParamNet.

\begin{figure}[t!]
    \centering
    \includegraphics[width=1.0\linewidth]{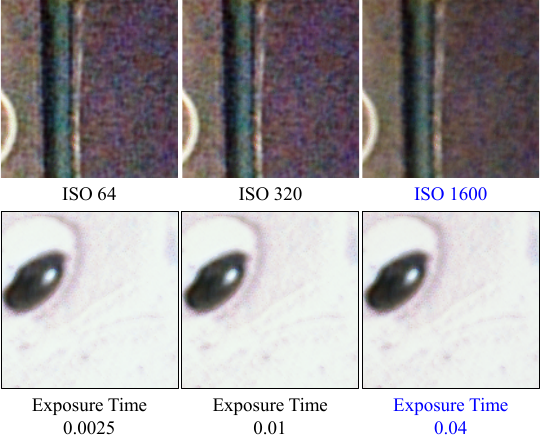}
    \vspace{-0.6cm}
    \caption{Example of sRGB reconstruction with modified optical parameters (sensitivity and exposure time) using the S7 ISP dataset~\cite{schwartz2018deepisp}. The blue text represents the ground-truth value.}
    \vspace{-0.4cm}
    \label{fig:manip-opt}
\end{figure}

\parhead{Optimal Performance of Each Module}
While ParamISP features a modularized network architecture where each module has specific objectives, all the modules are jointly trained in an end-to-end fashion utilizing a reconstruction loss.
Instead, we may explicitly train each module to serve its intended purpose.
Here, we compare these two training approaches.

Given that ParamNet takes only optical parameters as input, and that GlobalNet performs only global adjustment operations, it is clear that ParamNet and GlobalNet are trained to serve their respective purposes.
On the other hand, LocalNet may be trained to perform global operations as well as local ones.
To explicitly train LocalNet and GlobalNet for their respective purposes, we may 1) train GlobalNet without LocalNet, and 2) fix GlobalNet and train LocalNet.
We found that this sequential training results in a RAW reconstruction performance of 33.62dB, which is almost the same as that of the joint training (33.66dB).
For clarity, ParamNet was excluded from this experiment. The results suggest that although each module could be explicitly trained for its specific function, our joint training approach suffices to effectively train ParamISP.

\section{ParamNet \& Optical Parameters}
\label{sec:opticalparams-supp}
In Sec.~4.1 of the main paper, we analyze the impact of each optical parameter on performance in Tab.~2.
To supplement the experimental results in the main paper, we present additional qualitative and quantitative analyses here.

\begin{table}[t]
    \centering
    \scalebox{0.8}{
    \begin{tabularx}{1.12\columnwidth}{cccccc}
    \Xhline{2.5\arrayrulewidth}
    Otical Params & $A$ & $B$ & $C$ & $D$ & GT \\ \hline
    PSNR$\uparrow$ & 34.25 & 35.53 & 35.59 & \snd{35.99} & \fst{36.21} \\  
    SSIM$\uparrow$ & 0.9636 & 0.9682 & 0.9703 & \snd{0.9709} & \fst{0.9724} \\           
    \Xhline{2.5\arrayrulewidth}
    \end{tabularx}
    }
    \vspace{-0.2cm}
    \caption{Reconstruction quality of ParamISP with respect to varying optical parameters. $A$, $B$, $C$, and $D$ represent incorrect sensitivity, exposure time, aperture size, and focal length, respectively. GT represents ground-truth. ParamISP achieves the best reconstruction quality for ground-truth optical parameters, while the quality degrades for incorrect parameters, indicating that ParamISP can correctly reflect the adaptive behavior of real-world ISPs.}
    \vspace{-0.4cm}
    \label{tab:manip-opterror}
\end{table}

We first verify whether ParamNet operates to reflect the characteristics of optical parameters.
As shown in \cref{fig:manip-opt}, we manipulate the values of optical parameters and observe changes in visual results.  
Note that increasing the exposure-related optical parameters of ParamNet, such as aperture size, is not equivalent to increasing the actual exposure.
For instance, a manipulated high ISO value for the ISP does not mean that the final processed image should be brightened.
Instead, it tells the ISP that its input RAW data is captured with a high ISO value, which results in more noise.
The ISP can then adapt its operations, such as increasing the denoising strength.
ParamISP operates in the same way as well.
As the ISO value increases, the strength of denoising increases (1st row), and as the exposure time increases, saturated areas are restored more effectively (2nd row).
This experimental result shows that ParamISP can precisely mimic real-world ISP operations that change according to the optical parameters.

We also present a quantitative analysis.
In this analysis, we measure the reconstruction performance of ParamISP against various optical parameters. If ParamISP can accurately replicate the adaptive behavior of a real-world ISP, it is expected to attain optimal reconstruction quality with the ground-truth optical parameters, while the quality should diminish with incorrect parameters.
To this end, we use the D7000 test images of the RAISE dataset~\cite{dang2015raise} and their optical parameters.
Specifically, for each optical parameter of an image, we randomly change its value and measure its reconstruction quality. 
\cref{tab:manip-opterror} shows a result.
In the table, significant performance drops are observed when incorrect values are used for optical parameters.
This result indicates that ParamISP can correctly mimic the adaptive behavior of real-world ISPs.

\begin{figure}[t!]
    \centering
    \includegraphics[width=1.0\linewidth]{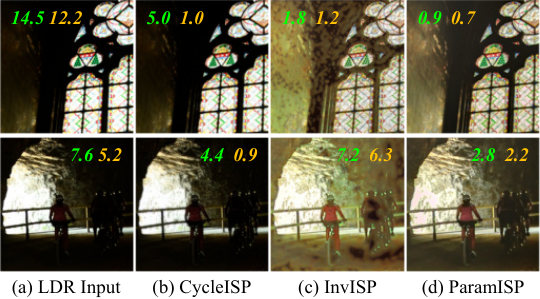}
    \vspace{-0.6cm}
     \caption{HDR reconstruction results of over-exposed regions. 
     Green: proportion of the pixels with values $\ge250$. 
     Yellow: proportion of the pixels with values $=255$. 
     The results show that our method produces a higher-quality HDR image with more details and also demonstrates a significant reduction in overly bright areas numerically.}
    \label{fig:hdr-supp2}
\end{figure}

\section{Additional Results}
\label{sec:moreresults-supp}

\begin{figure}[t!]
    \centering
    \includegraphics[width=1.0\linewidth]{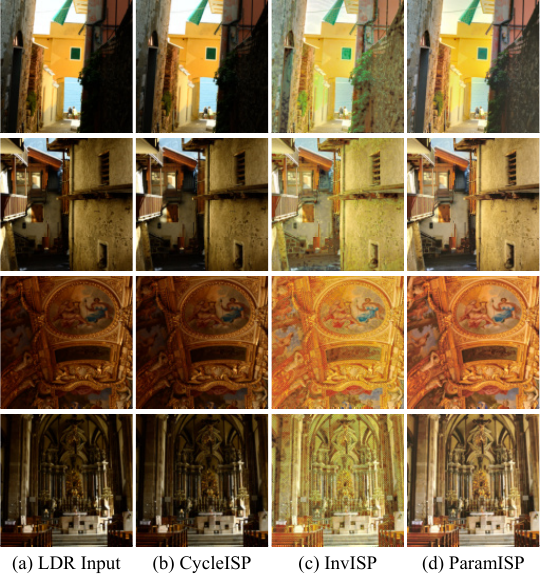}
    \vspace{-0.6cm}
    \caption{Qualitative examples of HDR reconstruction.}
    \label{fig:hdr-supp1}
\end{figure}

\begin{figure*}[p]
\includegraphics[width=1\linewidth]{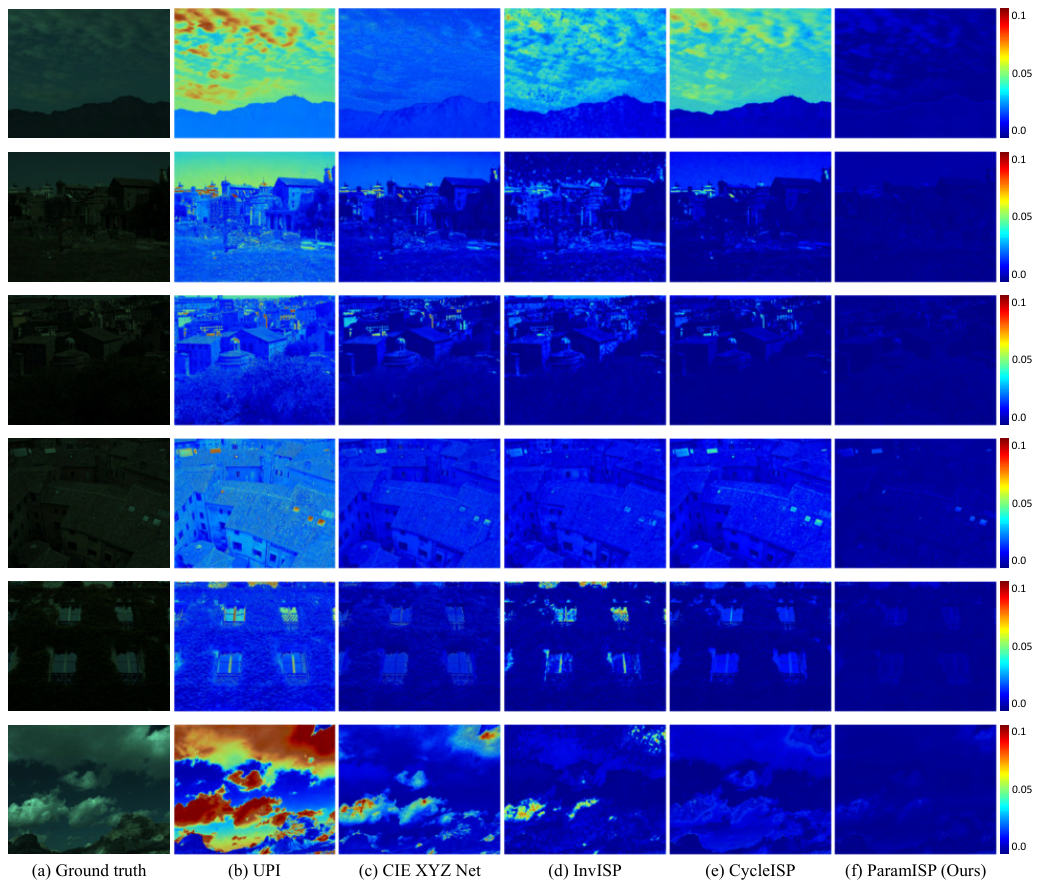}
\vspace{-0.2cm}
\caption{sRGB-to-RAW reconstruction. We show error maps between reconstructed and GT RAW images.}
\label{fig:inverse-supp}
\end{figure*}

\begin{figure*}[p]
\includegraphics[width=1\linewidth]{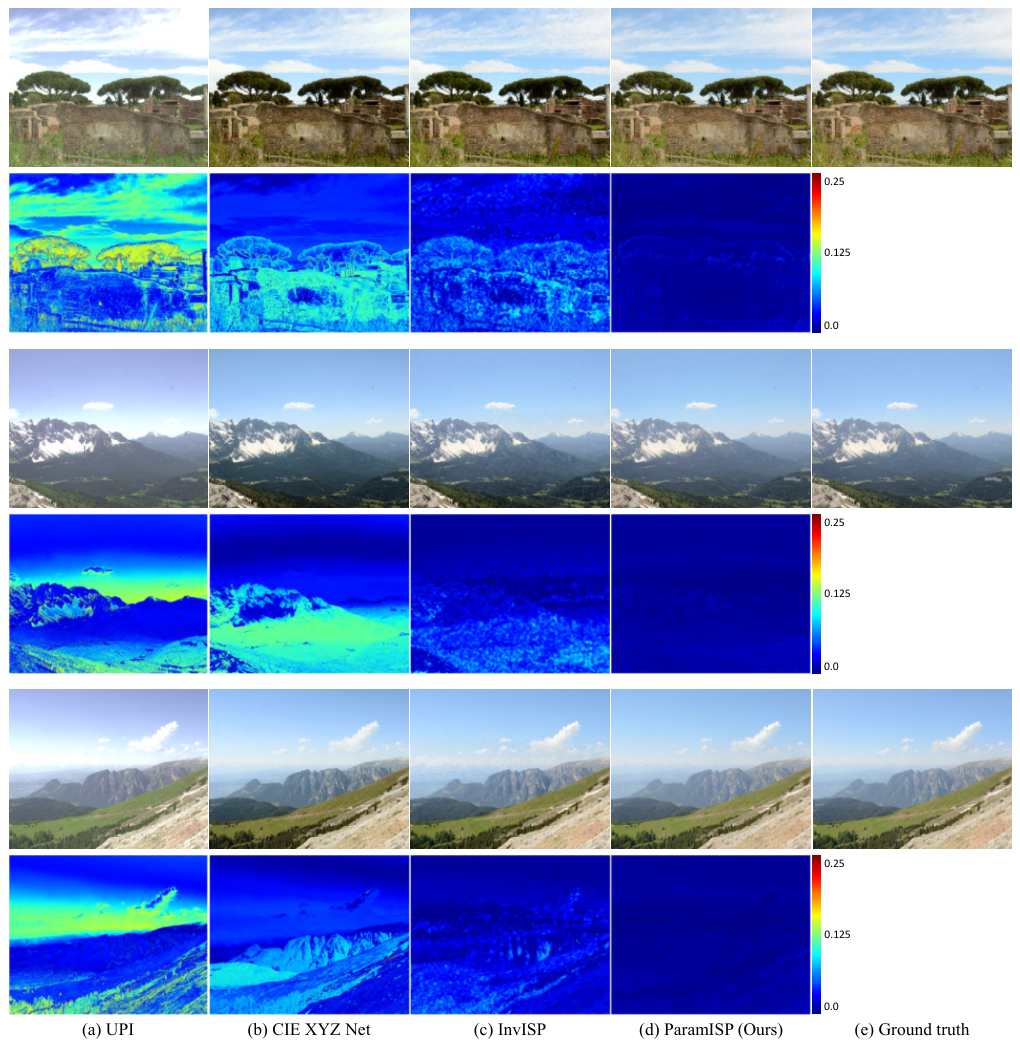}
\vspace{-0.2cm}
\caption{RAW-to-sRGB reconstruction. We show error maps between reconstructed and GT sRGB images.}
\label{fig:forward-supp}
\end{figure*}

We provide additional detailed quantitative results to supplement the experimental results in the main paper: an ablation study on the effects of input features and the proposed ParamNet (\cref{tab:abl-components}), the training strategy (\cref{tab:abl-strategy}), and comparison on RAW \& sRGB reconstruction (\cref{tab:recon-results}).
The first two results supplement the experimental results in Sec.~4.1 of the main paper, while the third one supplements Sec.~4.2.
Furthermore, we show additional qualitative results on HDR reconstruction (\cref{fig:hdr-supp2} \& \cref{fig:hdr-supp1}), sRGB-to-RAW reconstruction (\cref{fig:inverse-supp}), and RAW-to-sRGB reconstruction (\cref{fig:forward-supp}).
In all qualitative results, ParamISP shows visually better results compared to other methods, confirming its superior performance.

\begin{figure}[t!]
    \centering
    \includegraphics[width=1.0\linewidth]{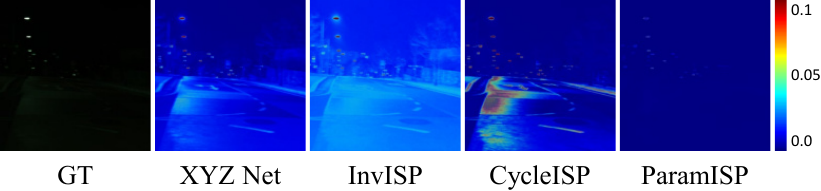}
    \vspace{-0.6cm}
     \caption{Qualitative results obtained by performing sRGB-to-RAW reconstruction using official pretrained models.
     We show error maps between the reconstructed and ground-truth (GT) RAW images.
     The GT RAW image in this figure is demosaicked for visualization.
     }
    \vspace{+0.4cm}
    \label{fig:offical-supp}
\end{figure}

We also report a quantitative comparison (sRGB-to-RAW) using the official pretrained models provided by the authors on the Sony A7R3 dataset~\cite{rim2020real} (Ours: 48.33 (dB), CIE XYZ Net: 27.86, InvISP: 26.43, CycleISP: 25.20).
It is worth noting that the official pretrained models of the previous methods were trained on different cameras than the target camera, which explains their lower performance.
\cref{fig:offical-supp} shows qualitative results. 
Our model, trained and evaluated on the Sony A7R3 dataset, demonstrates minimal reconstruction errors.
This indicates the importance of training neural ISP models on the target camera, as ISPs are dependent on the specific characteristics of cameras.

\section{Network Architecture}
\label{sec:networkarch-supp}
\begin{figure*}[p]
\includegraphics[width=1\linewidth]{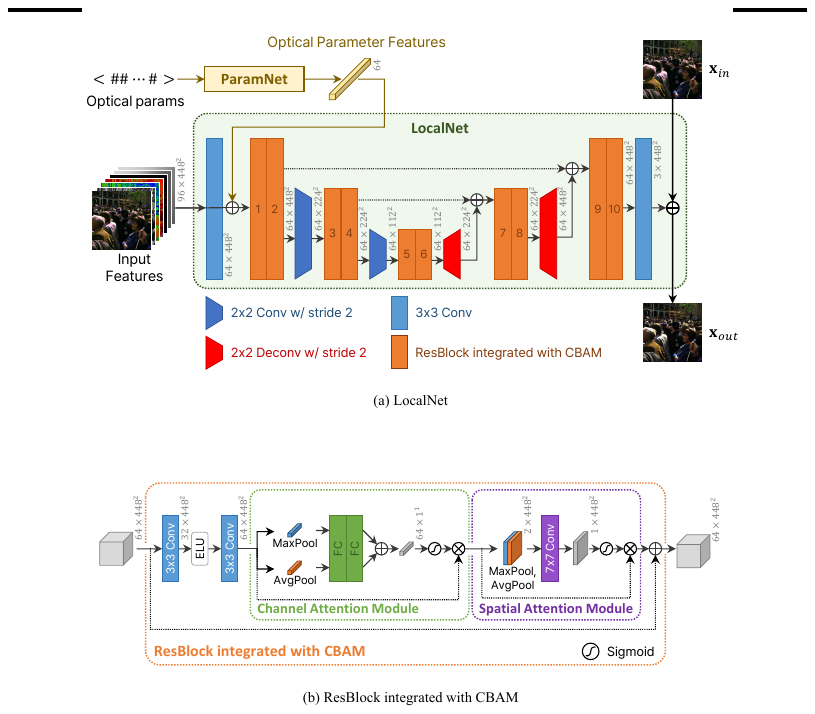}
\vspace{-0.2cm}
\caption{Detailed architecture of LocalNet.}
\label{fig:local-supp}
\end{figure*}

\begin{figure*}[t]
\includegraphics[width=1\linewidth]{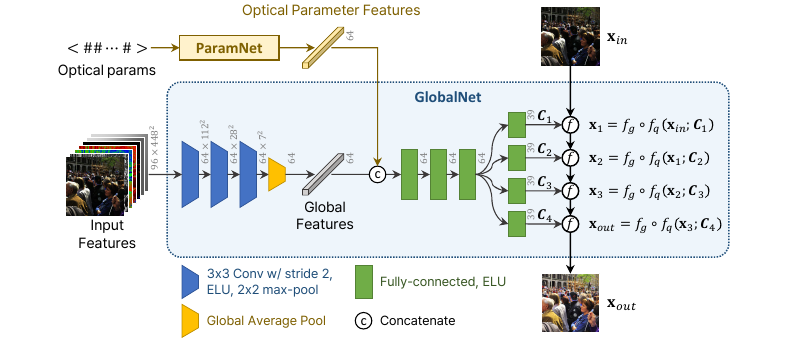}
\vspace{-0.2cm}
\caption{Detailed architecture of GlobalNet.}
\label{fig:global-supp}
\end{figure*}

\begin{figure*}[t]
\includegraphics[width=1\linewidth]{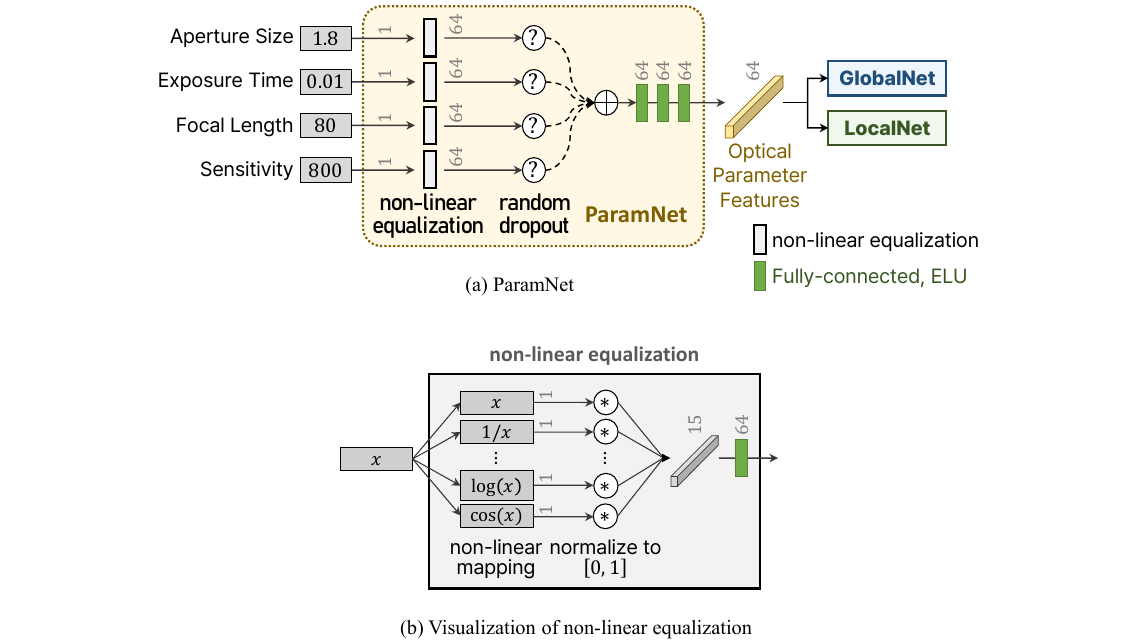}
\vspace{-0.2cm}
\caption{Detailed architecture of ParamNet.}
\label{fig:paramnet-supp}
\end{figure*}

We visualize detailed network architectures for LocalNet (\cref{fig:local-supp}), GlobalNet (\cref{fig:global-supp}), and ParamNet (\cref{fig:paramnet-supp}).

\begin{table*}[p]
\aboverulesep = -0.2mm
\belowrulesep =  0.1mm
    \centering
    \scalebox{0.75}{
    \begin{tabularx}{1.355\linewidth}{ll *{10}c *{2}{>{\columncolor{gray!20}}c}}
    \Xhline{2.5\arrayrulewidth}
    & \multirow{2}{*}{Components}
    & \multicolumn{2}{c}{D7000~\cite{dang2015raise}}
    & \multicolumn{2}{c}{D90~\cite{dang2015raise}}
    & \multicolumn{2}{c}{D40~\cite{dang2015raise}}
    & \multicolumn{2}{c}{S7~\cite{schwartz2018deepisp}}
    & \multicolumn{2}{c}{A7R3~\cite{rim2020real}}
    & \multicolumn{2}{c}{\cellcolor{gray!20}\textbf{Average}}\\
    \cmidrule(lr){3-4}
    \cmidrule(lr){5-6}
    \cmidrule(lr){7-8}
    \cmidrule(lr){9-10}
    \cmidrule(lr){11-12}
    \cmidrule(lr){13-14}
    &&      PSNR  &        SSIM
    &       PSNR  &        SSIM
    &       PSNR  &        SSIM
    &       PSNR  &        SSIM
    &       PSNR  &        SSIM 
    &       PSNR  &        SSIM  \\
    \midrule \multirow{4}{*}{\makecell{sRGB \\ → RAW}}
    & \!\!Ours (w/o Input Features \& ParamNet)
    &      33.66  &      0.9646
    &      34.50  &      0.9551
    &      44.87  & \snd{0.9879}
    &      34.51  & \snd{0.9074}
    &      44.66  &      0.9892
    &      38.44  &      0.9608  \\
    &\,\,\,\,\,\,\,\,\,\,\,\,$+$Input Features
    &      34.77  & \snd{0.9712}
    &      34.98  &      0.9672
    &      44.95  &      0.9847
    &      34.45  &      0.9063
    &      46.72  &      0.9912
    &      39.17  &      0.9641  \\
    \arrayrulecolor{gray} \cmidrule(l{0pt}r{0pt}){2-14} \arrayrulecolor{black}
    &\,\,\,\,\,\,\,\,\,\,\,\,$+$ParamNet w/o dropout
    & \snd{35.64} &      0.9702
    & \snd{36.26} & \snd{0.9724}
    & \snd{45.41} &      0.9858
    & \snd{34.81} &      0.9007
    & \snd{47.47} & \snd{0.9922}
    & \snd{39.92} & \snd{0.9643}  \\
    &\,\,\,\,\,\,\,\,\,\,\,\,\,\,\,\,\,\,\,\,\,\,\,\,$+$w/ dropout
    & \fst{36.21} & \fst{0.9724}
    & \fst{36.31} & \fst{0.9731}
    & \fst{45.73} & \fst{0.9883}
    & \fst{35.14} & \fst{0.9115}
    & \fst{47.80} & \fst{0.9922}
    & \fst{40.24} & \fst{0.9675}  \\
    \midrule\midrule \multirow{4}{*}{\makecell{RAW \\ → sRGB}}
    &\!\!Ours (w/o Input Features \& ParamNet)
    &      29.12  &      0.9399
    &      29.46  &      0.9544
    &      38.81  & \snd{0.9843}
    & \snd{28.48}  &      0.7593
    &      44.46  &      0.9806
    &      34.07  &      0.9237  \\
    &\,\,\,\,\,\,\,\,\,\,\,\,$+$Input Features
    &      29.21  &      0.9381
    &      29.49  &      0.9558
    &      39.08  &      0.9840
    &      28.42  &      0.7635
    &      44.39  &      0.9805 
    &      34.12  &      0.9244  \\
    \arrayrulecolor{gray} \cmidrule(l{0pt}r{0pt}){2-14} \arrayrulecolor{black}
    &\,\,\,\,\,\,\,\,\,\,\,\,$+$ParamNet w/o dropout
    & \snd{29.87} & \snd{0.9421}
    & \snd{30.39} & \snd{0.9639}
    & \snd{39.11} &      0.9836
    &      28.43  & \snd{0.7668}
    & \snd{44.63} & \snd{0.9808} 
    & \snd{34.49} & \snd{0.9274}  \\
    &\,\,\,\,\,\,\,\,\,\,\,\,\,\,\,\,\,\,\,\,\,\,\,\,$+$w/ dropout
    & \fst{29.89} & \fst{0.9422}
    & \fst{30.50} & \fst{0.9664}
    & \fst{39.63} & \fst{0.9849}
    & \fst{28.60} & \fst{0.7690}
    & \fst{44.78} & \fst{0.9815} 
    & \fst{34.68} & \fst{0.9288} \\
    \Xhline{2.5\arrayrulewidth}
    \end{tabularx}
    }
    \vspace{-0.2cm}
    \caption{Ablation study on the effects of input features and the proposed ParamNet. Each result was trained and evaluated using a specific camera only.}
    \vspace{16pt}
    \label{tab:abl-components}

    \centering
    \scalebox{0.8}{
    \begin{tabularx}{1.228\linewidth}{ll*{10}c *{2}{>{\columncolor{gray!20}}c}}
    \Xhline{2.5\arrayrulewidth}
    \multirow{2}{*}{} &
    \multirow{2}{*}{Strategy}
    & \multicolumn{2}{c}{D7000~\cite{dang2015raise}}
    & \multicolumn{2}{c}{D90~\cite{dang2015raise}}
    & \multicolumn{2}{c}{D40~\cite{dang2015raise}}
    & \multicolumn{2}{c}{S7~\cite{schwartz2018deepisp}}
    & \multicolumn{2}{c}{A7R3~\cite{rim2020real}}
    & \multicolumn{2}{c}{\cellcolor{gray!20}\textbf{Average}}\\
    \cmidrule(lr){3-4}
    \cmidrule(lr){5-6}
    \cmidrule(lr){7-8}
    \cmidrule(lr){9-10}
    \cmidrule(lr){11-12}
    \cmidrule(lr){13-14}
    &&      PSNR  &        SSIM
    &       PSNR  &        SSIM
    &       PSNR  &        SSIM
    &       PSNR  &        SSIM
    &       PSNR  &        SSIM
    &       PSNR  &        SSIM \\
    \midrule \multirow{4}{*}{\makecell{sRGB \\ → RAW}}
    & Generic
    &       34.62 &      0.9689
    &       35.61 &      0.9738
    &       41.37 &      0.9796
    &       34.65 &      0.9082
    &       44.04 &      0.9871
    &       38.06 &      0.9635 \\
    & Individual
    &      34.77  &      0.9712
    &      34.98  &      0.9672
    &      44.95  &      0.9847
    &      34.45  &      0.9063
    &      46.72  &      0.9912
    &      39.17  &      0.9641 \\
    & Generic$+$individual
    & \snd{36.47} & \snd{0.9758}
    & \snd{36.59} & \snd{0.9796}
    & \snd{45.75} & \fst{0.9879}
    & \snd{34.99} & \snd{0.9118}
    & \snd{47.18} & \snd{0.9920} 
    & \snd{40.20} & \snd{0.9694} \\
    \arrayrulecolor{gray} \cmidrule(l{0pt}r{0pt}){2-14} \arrayrulecolor{black}
    & ParamISP (Gen + Ind)
    & \fst{38.49} & \fst{0.9809}
    & \fst{37.06} & \fst{0.9810}
    & \fst{45.97} & \snd{0.9877}
    & \fst{35.20} & \fst{0.9125}
    & \fst{48.33} & \fst{0.9930}
    & \fst{41.01} & \fst{0.9710} \\
    \midrule\midrule \multirow{4}{*}{\makecell{RAW \\ → sRGB}}
    & Generic
    &      28.71  &      0.9262
    &      28.44  &      0.9447
    &      34.90  &      0.9685
    &      28.06  &      0.7580
    &      39.51  &      0.9603
    &      31.92  &      0.9115 \\
    & Individual
    &      29.21  &      0.9381
    &      29.49  & \snd{0.9558}
    & \snd{39.08} & \snd{0.9840}
    &      28.42  &      0.7635
    &      44.39  &      0.9805
    &      34.12  &      0.9244 \\
    & Generic$+$individual
    & \snd{31.51} & \snd{0.9491}
    & \snd{29.50} &      0.9535
    &      38.97  &      0.9831
    & \snd{28.42} & \snd{0.7716}
    & \snd{44.95} & \snd{0.9823} 
    & \snd{34.67} & \snd{0.9279} \\
    \arrayrulecolor{gray} \cmidrule(l{0pt}r{0pt}){2-14} \arrayrulecolor{black}
    & ParamISP (Gen + Ind)
    & \fst{34.14} & \fst{0.9628}
    & \fst{30.83} & \fst{0.9670}
    & \fst{39.54}  & \fst{0.9844}
    & \fst{29.02} & \fst{0.7868}
    & \fst{45.51}  &\fst{0.9841} 
    & \fst{35.81} & \fst{0.9370} \\
    \Xhline{2.5\arrayrulewidth}
    \end{tabularx}
    }
    \vspace{-0.2cm}
    \caption{Ablation study on the effects of the training strategy.
    `Generic', `Individual', and `Generic$+$individual' mean models trained using multiple camera datasets, models trained using only target camera datasets, and models trained using our two-stage training scheme, respectively. All the `Generic', `Individual' and `Generic$+$individual' models are without ParamNet.
    We also include our final model `ParamISP' trained using our two-stage training scheme in the table.}
    \vspace{16pt}
    \label{tab:abl-strategy}

    \centering
    \scalebox{0.8}{
    \begin{tabularx}{1.189\linewidth}{ll *{10}c *{2}{>{\columncolor{gray!20}}c}}
    \Xhline{2.5\arrayrulewidth}
    \multirow{2}{*}{} &
    \multirow{2}{*}{Method}
    & \multicolumn{2}{c}{D7000~\cite{dang2015raise}}
    & \multicolumn{2}{c}{D90~\cite{dang2015raise}}
    & \multicolumn{2}{c}{D40~\cite{dang2015raise}}
    & \multicolumn{2}{c}{S7~\cite{schwartz2018deepisp}}
    & \multicolumn{2}{c}{A7R3~\cite{rim2020real}}
    & \multicolumn{2}{c}{\cellcolor{gray!20}\textbf{Average}}\\
    \cmidrule(lr){3-4}
    \cmidrule(lr){5-6}
    \cmidrule(lr){7-8}
    \cmidrule(lr){9-10}
    \cmidrule(lr){11-12}
    \cmidrule(lr){13-14}
    &&      PSNR  &        SSIM
    &       PSNR  &        SSIM
    &       PSNR  &        SSIM
    &       PSNR  &        SSIM
    &       PSNR  &        SSIM
    &       PSNR  &        SSIM \\
    \midrule \multirow{5}{*}{\makecell{sRGB \\ → RAW}}
    & UPI~\cite{brooks2019unprocessing}
    &      20.67  &      0.7854
    &      26.57  &      0.8623
    &      22.05  &      0.7679
    &      29.98  &      0.8482
    &      30.48  &      0.9368
    &      25.95  &      0.8401 \\
    & CIE XYZ Net~\cite{afifi2021cie}
    &      30.04  &      0.9461
    &      32.62  &      0.9521
    &      38.57  &      0.9809
    &      33.24  &      0.8918
    &      36.42  &      0.9779
    &      34.18  &      0.9498 \\
    & CycleISP~\cite{zamir2020cycleisp}
    & \snd{35.52} & \snd{0.9740}
    & \snd{35.85} & \snd{0.9786}
    &      42.83  &      0.9831
    & \snd{34.55} &      0.9056
    &      45.35  &      0.9916
    &      38.82  & \snd{0.9666} \\
    & InvISP~\cite{xing2021invertible}
    &      33.48  &      0.9685
    &      35.39  &      0.9747
    & \snd{45.08} & \snd{0.9866}
    &      34.29  & \snd{0.9095}
    & \snd{47.14} & \snd{0.9924} 
    & \snd{39.08} &      0.9663 \\
    & ParamISP (Ours)
    & \fst{38.49} & \fst{0.9809}
    & \fst{37.06} & \fst{0.9810}
    & \fst{45.97} & \fst{0.9877}
    & \fst{35.20} & \fst{0.9125}
    & \fst{48.33} & \fst{0.9930} 
    & \fst{41.01} & \fst{0.9710} \\
    \midrule\midrule \multirow{4}{*}{\makecell{RAW \\ → sRGB}}
    & UPI~\cite{brooks2019unprocessing}
    &      18.81  &      0.6326
    &      20.30  &      0.8010
    &      16.01  &      0.7649
    &      20.05  &      0.4205
    &      19.37  &      0.5324 
    &      18.91  &      0.6303 \\
    & CIE XYZ Net~\cite{afifi2021cie}
    &      26.76  &      0.8703
    &      27.61  &      0.9183
    &      34.84  &      0.9635
    &      27.63  &      0.6978
    &      37.19  &      0.9396
    &      30.81  &      0.8779 \\
    & InvISP~\cite{xing2021invertible}
    & \snd{30.20} & \snd{0.9393}
    & \snd{28.89} & \snd{0.9448}
    & \snd{37.86} & \snd{0.9816}
    & \snd{28.96} & \snd{0.7862}
    & \snd{43.93} & \snd{0.9786} 
    & \snd{33.97} & \snd{0.9261} \\
    & ParamISP (Ours)
    & \fst{34.14} & \fst{0.9628}
    & \fst{30.83} & \fst{0.9670}
    & \fst{39.54} & \fst{0.9844}
    & \fst{29.02} & \fst{0.7868}
    & \fst{45.51} & \fst{0.9841} 
    & \fst{35.81} & \fst{0.9370} \\
    \bottomrule
    \end{tabularx}
    }
    \vspace{-0.2cm}
    \caption{Quantitative comparison on RAW \& sRGB reconstruction.
    Note that CycleISP is not included in this comparison because it needs an input sRGB image for sRGB reconstruction.}
    \label{tab:recon-results}
\end{table*}

{
    \small
    \bibliographystyle{ieeenat_fullname}
    \bibliography{reference_supp}
}
